\documentclass[a4paper,11pt]{article}
\usepackage{jheppub} 
\usepackage{amsthm}
\usepackage{tikz}

\newtheorem{theorem}{Theorem}

\makeatletter
\def\@fpheader{}
\makeatother

\title{\boldmath Counting AdS Vacua}

\author[1]{Zihni Kaan Baykara}
\author[2]{Alessandro Tomasiello}
\author[1]{Cumrun Vafa}
\affiliation[1]{Jefferson Physical Laboratory, Harvard University,\\
Cambridge, MA 02138, USA}
\affiliation[2]{Dipartimento di Matematica,\\
Università degli Studi di Milano-Bicocca,\\
Via Cozzi 55, 20126 Milano, Italy\\
and INFN, sezione di Milano-Bicocca}

\emailAdd{zbaykara@g.harvard.edu, alessandro.tomasiello@unimib.it, vafa@g.harvard.edu}

\abstract{We study the `number' $\mathfrak{N}(\mu)$ of AdS vacua with a UV cut off $ \mu$.  It has been proposed that this number is finite.  We find evidence that $\mathfrak{N}(\mu)\sim a \ \mu^{-b}$ as $\mu \rightarrow 0$ for some constants $a$ and $b$ of $O(1)$ in Planck units that may depend on dimension and the number of supercharges.  For this result to hold it is crucial to integrate over the volume of massless and tachyonic directions of AdS which corresponds to the volume of the space of marginal and relevant deformations of the dual CFT. We are led to the surprising prediction that theories with large number of light moduli contribute very little to the volume measure among all theories. We also speculate about the dS case leading to the number of quasi-dS vacua of the order of $\Lambda^{-\alpha}$ for some $O(1)$ parameter $\alpha$.}

\begin{document}
\maketitle
\flushbottom

\section{Introduction}
One of the main motivations of the Swampland program \cite{Vafa:2005ui} is the observation of the apparent finiteness of the number of quantum gravity vacua.  It is this finiteness which stands in sharp contrast to the naive expectation that there are infinitely many possible consistent theories of quantum gravity  motivated from naive expectations based on EFT arguments.  

This finiteness needs to be qualified, as families of supersymmetric solutions coming from string theory come in a family parameterized by the vev of the massless scalar fields, called the moduli space.  What one would have to mean by finiteness is presumably the total volume of such spaces.
However, this space has sometimes infinite volume.  Moreover, the number of AdS vacua is also naively infinite.  To fix these issues, it has been proposed that the notion of finiteness refers to the volume of theories with a fixed UV cutoff $\mu>0$  \cite{Acharya:2006zw,Hamada:2021yxy,Delgado:2024skw} (see also \cite{Grimm:2025lip,Grimm:2021vpn} for links to tame geometry).\footnote{We use $\mu$ to denote the cutoff to avoid confusion with the cosmological constant $\Lambda$ of AdS.} In particular, any light tower $m_{\mathrm{tower}}<\mu$ invalidates the EFT, and such weakly coupled towers arise at the infinite distances of the moduli space. Consequently, the space of vacua obeying the cutoff $\mu$ lie in the interior of the moduli space enclosing a finite volume $\mathfrak{N}(\mu)$, which corresponds to the number of such vacua. From the CFT perspective, related ideas about defining measures on the space of CFTs and requiring a finite gap above the vacuum to obtain a finite theory space were recently explored in \cite{Belin:2025qjm}.

However, this number could still diverge as $\mu\rightarrow 0$.
The dependence of the `number of vacua', including the volume of massless modes, to $\mu$ in the case of Minkowski vacua has been studied in \cite{Delgado:2024skw,Ooguri:2006in}.  One expects the number of Minkowski vacua to go as 
\begin{align}
\mathfrak{N}_{\text{Mink}}(\mu)\lesssim a|\log\mu|^b,
\end{align}
as $\mu \rightarrow 0$, where $a,b$ could depend on the dimension $d$ and the number of supersymmetries.  The main aim of this paper is to generalize this to the case of AdS vacua.

The fact that for AdS we naively have infinitely many vacua labeled by a number $N$ (such as the number of branes leading to the corresponding AdS) is avoided by imposing a cutoff.  The reason is that large $N$ would correspond to small $|\Lambda|\rightarrow 0$ and there is typically a KK tower of light states whose scale goes as $|\Lambda|^{1/2}$ which thus is bounded if we fix a cutoff $\mu$, by $|\Lambda|^{1/2}\gtrsim \mu$. We argue that this leads to power-law growth for the number of AdS vacua
\begin{align}
    \mathfrak{N}_{\mathrm{AdS}}(\mu)\sim a \mu^{-b}
\end{align}
as $\mu\rightarrow 0$.  We have to include all modes with mass less than $\mu$ in the EFT as $\mu \rightarrow 0$.  In particular, we have to include the volume of massless and tachyonic modes of AdS.  If we only include the volume of massless moduli and not the volume of tachyonic modes that correspond to flows between CFTs by RG flows, we find a counterexample to the above bound with $\mathfrak{N}_{\text{AdS}}(\mu)\sim \exp(A\mu^{-B})$. In the resolution of this puzzle we learn that theories which have many light moduli have smaller volume.
This surprising conclusion is related to the fact that the volume of a sphere of a fixed diameter decreases super-exponentially with dimension.

By the AdS/CFT correspondence, this conjecture leads to the statement that the count of the number of CFTs with a gravity dual and central charge less than $c$ also has power-law growth
\begin{align}
    \mathfrak{N}_{\text{CFT}}(c) \sim Ac^B.
\end{align}

The organization of this paper is as follows.  In \autoref{sec:Mink}, we review the count for the Minkowski vacua. In \autoref{sec:maxSUSY}, we do the count for AdS vacua with maximal SUSY. In \autoref{sec:lowSUSY}, we extend this to the cases with less supersymmetry and find it is important to include tachyonic modes.  In \autoref{sec:anthropic}, we end with some concluding thoughts in the context of de Sitter and in particular with connections to the anthropic principle. 

The supplementary material consists of a Mathematica notebook containing the full derivations and three appendices: in \autoref{app:instanton} we present the instanton‐restricted counts, in \autoref{app:Grassmannian} we review the relevant Grassmannian geometry, and in \autoref{app:Motzkin} we summarize the reduction method for systems of inequalities.

\section{Minkowski Vacua}\label{sec:Mink}

We now review the finiteness argument in Minkowski compactifications, following \cite{Delgado:2024skw}.
This setting isolates the contribution to the count from the moduli.

Consider a $d$-dimensional Minkowski compactification of string theory on an internal manifold $X$ with continuous moduli $\phi^i$.
The scalar kinetic term in the low-energy effective action defines a Riemannian metric $G_{ij}(\phi)$ on the moduli space $\mathcal{M}$,
\begin{align}
S_{\text{kin}} = \int \mathrm{d}^d x \sqrt{-g}\
G_{ij}(\phi) \partial_\mu \phi^i \partial^\mu \phi^j.
\end{align}
The associated measure is
\begin{align}
\mathrm{d}\nu_{\text{moduli}} = \sqrt{\det G}\ \mathrm{d}^n\phi,
\end{align}
where $n=\dim \mathcal{M}$.
We define the truncated region $\mathcal{M}_\mu$ as the subspace of moduli where the effective field theory remains valid, i.e.
\begin{align}
\mathcal{M}_\mu = \{\phi\in\mathcal{M}\mid m_{\mathrm{tower}}(\phi)>\mu\},
\end{align}
and denote its total moduli-space volume by
\begin{align}
\mathfrak N_{\mathrm{Mink}}(\mu) =V(\mathcal M_\mu)= \int_{\mathcal{M}_\mu} \mathrm{d}\nu_{\text{moduli}}.
\end{align}

The distance conjecture states that if we consider a region of diameter $L$ centered near a point in $\mathcal{M}$, a tower of states becomes exponentially light near the boundary of the region with characteristic mass scale:
\begin{align}
m_{\text{tower}}(L) \sim e^{-\alpha L},
\end{align}
where $\alpha = \mathcal{O}(1)$ and we write the expressions with $d$-dimensional Planck mass set to 1.
Fixing a cutoff $\mu$ therefore imposes a maximum radius
\begin{align}
L_{\max}(\mu) \sim \frac{1}{\alpha} |\log \mu|.
\end{align}
 The EFT is valid only inside the geodesic ball of radius $L_{\max}$ in moduli space.

As has been argued in \cite{Delgado:2024skw} the volume is either finite as $L\rightarrow \infty$ (due to dualities) or at most diverges like the Euclidean space (based on finiteness of the fully compactified Hilbert space).  This leads to the bound
\begin{align}
V(\mathcal{M}_\mu)
\lesssim L_{\max}^{n}
\sim |\log \hat\mu|^{n}.
\end{align}
  The bound currently realized in string theory is $n=2$ \cite{Delgado:2024skw}.
Physically, it means that the number of distinct EFT domains consistent up to cutoff $\mu$ grows at most polynomially in the logarithm of $\mu^{-1}$:
\begin{align}
\mathfrak{N}_{\mathrm{Mink}}(\mu) \lesssim |\log \mu|^{n},
\end{align}
with $n\sim O(1)$.

\section{Maximal SUSY AdS Vacua}\label{sec:maxSUSY}
Maximal supersymmetry arises from M-theory or string theory on AdS$_d\times S^{D-d}$ with Freund--Rubin flux. We do the analysis for general $d$ and $D$ in this section, but the actual cases with maximal supersymmetry are $d=4,7$ in M-theory ($D=11$) and $d=5$ in type IIB string theory ($D=10$).

In the maximal supersymmetry setting there are no supersymmetry-preserving moduli or  tachyons, and the only discrete datum specifying the vacuum is the flux integer $N\in\mathbb Z_{>0}$. Hence the counting reduces to summing over $N$ consistent with the cutoff.

Since we have two different dimensions, we have two different notions of Planck mass.
The count of vacua $\mathfrak{N}$ is a dimensionless number and is more natural to only use the lower dimensional EFT to measure the Planck mass.  So we introduce the dimensionless cutoff
\begin{align}
  \hat\mu \equiv \frac{\mu}{M_d}, 
\end{align}
where we denote the $d$-dimensional Planck mass by $M_d$. Note that the dimensionless cutoff takes values 
$\hat\mu\in(0,1)$, and our interest is the $\hat\mu \ll 1$ asymptotic regime.

\paragraph{Scales.}
Let $R$ be the radius of $S^{D-d}$. Flux quantization and dimensional reduction give
\begin{align}
    M_{D}^{D-d-1}\!\int_{S^{D-d}}\!F_{D-d}  \sim  M_{D}^{D-d-1} R^{D-d-1}  \sim  N, \qquad
  M_d^{d-2}  \sim  M_{D}^{D-2} R^{D-d},
\end{align}
from which we get useful dimensionless relationships
\begin{align}
  M_{D}R \sim N^{\frac{1}{D-d-1}}, \qquad (M_d R)^{d-2} \sim N^{\frac{D-2}{D-d-1}}, \qquad \frac{M_d}{M_D} \sim N^{\frac{D-d}{(d-2)(D-d-1)}}.
\end{align}
In IIB, the dilaton also appears in these relations, but we will not focus on it\footnote{Including such massless moduli can in principle affect our count by an extra factor of $|\log\mu|^a$.} and set it to a fixed value, of order 1.

\paragraph{Towers.} Even an unstable light tower is enough to invalidate the EFT. Therefore, we will be as comprehensive as possible in our list of possible towers in the proceeding analyses, including BPS as well as non-BPS towers. For example, although branes wrapped on trivial $q$-cycles on the equator of $S^p$ are unstable, they still form a tower. One would see their signature as resonances in the EFT amplitudes even though they would not directly show up in the spectrum as stable states.

In the maximal supersymmetric case, it turns out the lightest tower is the $S^{D-d}$ Kaluza--Klein tower, which we demand to be greater than the cutoff
\begin{align}\label{eq:Sphere-KK}
  \frac{m_{\rm KK}}{M_d} \sim \frac{1}{M_d R} \sim \left(N^{-\frac{D-2}{d-2}}\right)^{\frac 1 {D-d-1}}\gtrsim\hat\mu.
\end{align}
Other potential towers are parametrically heavier at large $N$ as  wrapped objects on $S^{D-d}$ get heavier with increasing $N$ due to $S^{D-d}$ getting larger. 

More explicitly, a $p$-brane wrapped on a $q$-cycle of $S^{D-d}$ is a $(p-q)$-brane on spacetime with tension
\begin{align}
    T_{p-q} \sim M_{D}^{p+1} R^{q}.
\end{align}
The tower scale associated with a $(p-q)$-brane must satisfy
\begin{align}
    T^{\frac{1}{p-q+1}}_{p-q} \sim M_{D} (M_DR)^{\frac q {p-q+1}} \gtrsim \mu,
\end{align}
or in terms of the dimensionless cutoff,
\begin{align}
    \frac{M_{D}}{M_{d}} (M_{D} R)^{\frac q {p-q+1}} \sim \left(N^{\frac{q}{p-q+1}-\frac{D-d}{d-2}} \right)^{\frac 1 {D-d-1}} \gtrsim\hat \mu.\label{eq:Sphere-wrapped}
\end{align}
We see that since we consider $d>2$, the exponent of $N$ in \eqref{eq:Sphere-wrapped} is always greater than the exponent in \eqref{eq:Sphere-KK}. Therefore the lightest tower is KK.
\paragraph{AdS count.}
EFT validity requires $m_{\rm tower}>\mu$. Since $m_{\rm tower}=m_{\rm KK}$ is the smallest scale, the count is just the number of integers $N$ such that
\begin{align}
  1 \leq N \lesssim \hat\mu^{-\frac{(D-d-1)(d-2)}{D-2}}.
\end{align}
Therefore the final count is
\begin{align}\label{eq:count-maxSUSY}
  \mathfrak{N}_{AdS_d}^{(Q=16)}(\hat\mu)
 \sim \hat\mu^{-\frac{(D-d-1)(d-2)}{D-2}}.
\end{align}

\paragraph{CFT count.}
Equivalently, in terms of the CFT central charge $c\sim (M_d R)^{d-2}\sim N^{\frac{D-2}{D-d-1}}$, the inequality is
\begin{align}
    1\leq c \lesssim \hat\mu^{-(d-2)}.
\end{align}
The maximum central charge $\hat c$ is then
\begin{align}
    \hat c \equiv \hat \mu^{-(d-2)}.
\end{align}
For maximally supersymmetric case we have $c\sim N^{\frac{d-1}{2}}$ (i.e., $N^3,N^2,N^{\frac{3}{2}}$ for $d=7,5,4$ cases). Counting in terms of $N$, we find that the count of CFTs with a gravity dual and central charge less than $\hat c$ are given as
\begin{align}
    \mathfrak{N}_{\text{CFT}_{d-1}}^{(Q=16)}  (\hat c) \sim \hat c^{\frac{2}{d-1}}\sim {\hat \mu}^{\frac{-2(d-2)}{d-1}}.
\end{align}
This way of writing it is more useful as it only refers to the dimension $d$ of AdS and not its realization in string theory or M-theory, which requires in addition $D$ which is invisible to an EFT.

\paragraph{In terms of $\Lambda$.}
We can also write the count in terms of the cosmological constant $\Lambda$. We define the dimensionless cosmological constant
\begin{align}
    \hat \Lambda \equiv\frac{\Lambda}{M_d^2}.
\end{align}
Then, using $|\Lambda|\sim R^{-2}$, we have
\begin{align}
    |\hat\Lambda| \sim \hat\mu^2,
\end{align}
and the count is
\begin{align}
    \mathfrak{N}_{AdS_d}(\hat\Lambda)\sim |\hat\Lambda|^{-\frac{d-2}{d-1}}.
\end{align}
The final counts for various $d$ are summarized in \autoref{tab:max-susy}.
\begin{table}[ht]
    \centering
    \begin{tabular}{|c|c|c|c|}
    \hline
    \multicolumn{4}{|c|}{$Q=16$ Counts}\\
    \hline
        $\mathrm{AdS}_d$ &  $\mathfrak{N}_{AdS}(\hat\Lambda)\sim |\hat \Lambda|^{-\frac{d-2}{d-1}}$& $\mathfrak{N}_{AdS}(\hat\mu)\sim {\hat \mu}^{\frac{-2(d-2)}{d-1}}$& $\mathfrak{N}_{CFT}\sim {\hat c}^{\frac{2}{d-1}}$\\\hline\hline
         7 &  $|\hat\Lambda|^{-5/6}$ & $\hat\mu^{-5/3}$ & $\hat c^{1/3}$\\
         5 &  $|\hat\Lambda|^{-3/4}$ & $\hat\mu^{-3/2}$ & $\hat c^{1/2}$\\
         4 &  $|\hat\Lambda|^{-2/3}$ & $\hat\mu^{-4/3}$ & $\hat c^{2/3}$\\\hline
    \end{tabular}

    \caption{The counts $\mathfrak{N}$ of maximally supersymmetric $\mathrm{AdS}_d$ and the dual $CFT_{d-1}$ in terms of the dimensionless parameters $\hat \Lambda$, UV-cutoff parameter $\hat \mu\in (0,1)$ and the maximum central charge $\hat c$ for various dimensions $d$. }
    \label{tab:max-susy}
\end{table}

\section{Lower SUSY AdS Vacua}\label{sec:lowSUSY}

In theories with less than maximal supersymmetry, the count of AdS vacua involves both discrete and continuous ingredients. The discrete part comes from flux integers and orbifold data such as $(N, k_i)$ while the continuous part arises from the scalar manifold of the corresponding gauged supergravity, whose tachyonic and marginal directions must be integrated appropriately with the cutoff. Each AdS dimension presents distinct mechanisms known in string landscape for generating these vacua.

In the following we will analyze some classes of AdS vacua, which we think give a good illustration of the general case for the counting.
In \autoref{sec:AdS7}, we do the count in the simplest setting of AdS$_7\times S^4/\mathbb Z_k$ in M-theory and of AdS$_7\times M_3$ in IIA, and go through various schemes for incorporating the tachyonic modes in the count. This section is the technical crux of the paper, as the count factor involving the tachyons is a subtle issue. In \autoref{sec:AdS5}, we consider AdS$_5$ with sphere quotients, Sasaki-Einstein manifolds, as well as Class S theories. In \autoref{sec:AdS4}, we lastly consider sphere quotients of AdS$_4$.\footnote{One can also consider nonabelian orbifolds  but one expects those will not lead to much more than the abelian ones when doing sphere quotient orbifolds.}
There would be \emph{many} more classes of solutions to consider; we offer a bird's eye view in \autoref{sub:disc}. A conservative variant of the counts for the sphere quotients that imposes instanton-action constraints is presented in \autoref{app:instanton}. The complete derivations are worked out explicitly in the supplementary Mathematica notebook.

\subsection{\texorpdfstring{$\mathrm{AdS}_7$}{AdS7}}
\label{sec:AdS7}

There are two types of AdS$_7$ solutions: M-theory on a supersymmetric sphere quotient $\mathrm{AdS}_7 \times S^4/\mathbb Z_k$, with $N$ internal units of $F_4$; and IIA on AdS$_7\times M_3$, with $N$ internal units of $H$, as well as D8/D6-branes in various configurations \cite{Apruzzi:2013yva,Apruzzi:2015wna,Cremonesi:2015bld}.\footnote{Various variants are possible, such as including an $E_8$ wall in M-theory or O-planes in IIA; we don't expect these to add much the overall picture, and we will not consider them in what follows.} The latter class is related to the former: as a result of tachyon condensation, or (from the dual SCFT point of view) as Higgsing of flavor symmetries.

We will begin by counting the M-theory solutions; the issue of tachyonic modes will then lead us to a way to count the latter.

The count of all the M-theory vacua has two parts: a discrete count of the $(N,k)$ that obey the cutoff, and a volume factor $V_{N,k}(\hat \mu)$ for each discrete choice associated to the continuous degrees of freedom. The total is then given by
\begin{align}
    \mathfrak{N}_{\mathrm{AdS}_7}^{(Q=8)}(\hat\mu) = \sum_{N,k} V_{N,k}(\hat\mu).
\end{align}

\subsubsection{Discrete count}
\paragraph{Scales.}
Let $R$ be the radius of $S^{4}$. Flux quantization and dimensional reduction give
\begin{align}
    M_{11}^{3}\!\int_{S^{4}/\mathbb Z_k}\!F_{4}  \sim  M_{11}^{3} \frac{R^{3}}{k}  \sim  N, \qquad
  M_7^{5}  \sim  M_{11}^{9} \frac{R^{4}}{k},
\end{align}
from which we get useful dimensionless relationships
\begin{align}
  M_{11}R \sim (Nk)^{1/3}, \qquad (M_7 R)^{5} \sim N^3 k^2, \qquad \frac{M_7}{M_{11}} \sim N^{4/15}k^{1/15}.
\end{align}

\paragraph{Towers.} 

For the present case, it again turns out the lightest tower is the $S^4$ KK tower with scale
\begin{align}\label{eq:Sphere/Zk-KK}
  \frac{m_{\rm KK}}{M_7} \sim \frac{1}{M_7 R} \sim N^{-3/5}k^{-2/5}\gtrsim\hat\mu.
\end{align}

As mentioned in the previous section, even an unstable light tower is enough to invalidate the EFT. Therefore we must investigate all possibilities of $p$-branes wrapping $q$-cycles of $S^4/\mathbb Z_k$, even though some may be trivial cycles and thus unstable. In principle, for large $k$, the cycles of $S^4/\mathbb Z_k$ shrink, thus wrapped brane towers become light, but as we will now show they are always heavier than the KK scale for this case.

The tension of a $p$-brane wrapping a $q$-cycle of $S^4/\mathbb Z_k$ for $0< q\leq \min(p,4)$ is given by
\begin{align}
    T_{p-q}\sim M_{11}^{p+1} \frac{R^q}{k}. \label{eq:Sphere/Zk-tension}
\end{align}
This is because a $q$-cycle has volume $R^q$ in the $k$-sheeted covering $S^4$ of $S^4/\mathbb Z_k$. Therefore the volume of the $q$-cycle in the quotient is $R^q/k$. Lastly, for $q=0$ we simply have
\begin{align}
    T_p \sim M_{11}^{p+1}.
\end{align}

In terms of the dimensionless cutoff, the associated $(p-q)$-brane tower scale is
\begin{align}
    \frac{T_{p-q}^{\frac{1}{p-q+1}}}{M_7}\sim \frac{M_{11}}{M_{7}} \left(\frac{(M_{11} R)^q}{k}\right)^{\frac 1 {p-q+1}} \sim  N^{-\frac 4 {15} +\frac{q}{3(p-q+1)}}k^{-\frac{1}{15}+\frac{q-3}{3(p-q+1
    )}}\gtrsim \hat \mu.\label{eq:Sphere/Zk-wrapped}
\end{align}
Comparing the exponents of $N$ in the KK tower \eqref{eq:Sphere/Zk-KK} and the wrapped brane tower \eqref{eq:Sphere/Zk-wrapped},
\begin{align}
    -\frac 3 5 < -\frac 4 {15} + \frac{q}{3(p-q+1)},
\end{align}
since $p>q$. Similarly comparing the exponents of $k$ we have
\begin{align}
    -\frac 2 5 \leq -\frac{1}{15}+ \frac{q-3}{3(p-q+1)}.
\end{align}
since the only way the second term on the RHS can be negative is if $q=1$, for which neither M2 nor M5 with $p=2,5$ violate the claimed inequality. So wrapped M2 and M5 brane tower scales are no lighter than the KK tower scale, therefore we can ignore them.

Lastly, for $q=0$, we have
\begin{align}
    \frac{T_{p}^{\frac{1}{p+1}}}{M_7}\sim \frac{M_{11}}{M_{7}} \sim  N^{-\frac 4 {15}}k^{-\frac{1}{15}}\gtrsim \hat \mu,
\end{align}
which is also heavier than the KK tower scale. Thus, the KK tower has the lightest scale.

The analysis of scales and towers carried out so far can be repeated for the IIA solutions. The internal space $M_3$ is an $S^2$ fibered over an interval \cite{Apruzzi:2013yva,Apruzzi:2015wna,Cremonesi:2015bld}, with D8/D6-branes located at special points of the interval. $N$ is now the internal $H$ flux, and $k$ the maximum flux of $F_2$. ($\int F_2$ jumps when one crosses the branes.) There are several more states to consider in the tower analysis, but in the end we still have that the only relevant condition is \eqref{eq:Sphere/Zk-KK}. We will come back to these solutions soon.

\paragraph{Cylindrical decomposition.}
We have three bounds on $N$ and $k$: one coming from $m_{KK}>\mu$ and two from positivity of the flux $N$ and orbifold order $k$,
\begin{subequations}
\begin{align}
    N^3k^2 &\lesssim  \hat\mu^{-5},\\
    1&\leq N,\label{eq:N-positive}\\
    1 &\leq  k.
\end{align}
\end{subequations}
We now put the inequalities in a form that makes it easy to count the number of allowed $(N,k)$. 

We have a combined upper and lower bound for $k$ as
\begin{align}\label{eq:k-ineq}
    1 \leq k \lesssim \hat\mu^{-5/2}N^{-3/2}.
\end{align}
For this inequality to be consistent, the outer inequality must hold: $1 \lesssim \hat\mu^{-5/2}N^{-3/2}$, which when combined with \eqref{eq:N-positive} gives a lower and upper bound for $N$ without using $k$:
\begin{align}\label{eq:ads7-Nineq}
    1 \leq N \lesssim \hat\mu^{-5/3}.
\end{align}
Once $N$ is fixed to a legal value, the bound on $k$ is given by \eqref{eq:k-ineq}. Thus we obtained a streamlined list of inequalities. This method of reducing inequalities is known as \textit{Fourier--Motzkin elimination}, and more generally as \textit{cylindrical decomposition}.

\paragraph{AdS count.}
By the cylindrical decomposition, we count the number of $N$ and $k$ that satisfy \eqref{eq:k-ineq}, \eqref{eq:ads7-Nineq}.
The final count is
\begin{align}\label{eq:count-AdS7-lowSUSY}
    \mathfrak{N}_{\mathrm{AdS}_7}^{(Q=8)}(\hat\mu) \sim \sum_{N=1}^{\lfloor\hat\mu^{-5/3}\rfloor} \sum_{k=1}^{\lfloor\hat\mu^{-5/2} N^{-3/2}\rfloor} V_{N,k}(\hat\mu).
\end{align}

Note that we still did not specify the volume factor $V_{N,k}(\hat\mu)$ associated to a fixed $N$ and $k$. Naively, we might consider assigning a weight
\begin{align}\label{eq:V-naive}
    V_{N,k}(\hat\mu) \overset{\text{naive}}{=} 1
\end{align}
to each choice of $N$ and $k$, so that we are only doing a discrete count of the number of $(N,k)$ consistent with the cutoff. This assignment turns out to be correct after nontrivial and subtle considerations. We defer further discussion of the volume factor to the next section.

With the volume factor taken as $1$, we approximate the sum in \eqref{eq:count-AdS7-lowSUSY} by an integral
\begin{align}
    \mathfrak{N}_{\mathrm{AdS}_7}^{(Q=8)}(\hat\mu) &\sim \int_1^{\hat\mu^{-5/3}}dN\int_1^{\hat\mu^{-5/2}N^{-3/2}} dk.
\end{align}
Evaluating, we get the final count
\begin{align}
    \mathfrak{N}_{\mathrm{AdS}_7}^{(Q=8)}(\hat\mu) \sim \hat\mu^{-5/2}.
\end{align}
In terms of the cosmological constant $|\hat\Lambda| \sim \hat\mu^2$,
\begin{align}
    \sim |\hat\Lambda|^{-5/4}.
\end{align}
Note that the count is larger than the maximally supersymmetric case \eqref{eq:count-maxSUSY} for $d=7$ and $D=11$ as expected, since supersymmetry constrains the theory space. The same goes for the corresponding CFT count
\begin{align}
    \mathfrak{N}_{\mathrm{CFT}_6}^{(Q=8)}(\hat c) \sim \hat c^{1/2}.
\end{align}

\subsubsection{Volume factor}
\label{sub:vol}

\paragraph{The scalar manifold.}

To understand what the volume factor is for fixed compactification data $(N,k)$,
it is useful to review the classification of  
holographic six-dimensional $(1,0)$ SCFTs. 
We begin with the dual of M-theory on  
$\mathrm{AdS}_7\times S^4/\mathbb Z_k$, which is 
the theory of $N$ coincident M5 branes probing an 
$A_{k-1}$ singularity. 

It is a strongly coupled  
$6d$ $(1,0)$ SCFT whose tensor branch effective description is a linear quiver gauge theory:
\begin{center}
\begin{tikzpicture}[
  x=1cm,y=1cm,
  flavor/.style = {draw, rectangle, minimum size=7mm, thick},
  gauge/.style  = {draw, circle,    minimum size=7mm, thick}
]

\node[flavor, label=above:{$SU(k)$}] (FL) at (0,0) {};

\node[gauge,  label=above:{$SU(k)$}] (G1) at (2,0) {};
\node[gauge,  label=above:{$SU(k)$}] (G2) at (4,0) {};
\node            (Dots)              at (6,0) {$\cdots$};
\node[gauge,  label=above:{$SU(k)$}] (Gm) at (8,0) {};

\node[flavor, label=above:{$SU(k)$}] (FR) at (10,0) {};

\draw[thick] (FL) -- (G1);
\draw[thick] (G1) -- (G2);
\draw[thick] (G2) -- (Dots);
\draw[thick] (Dots) -- (Gm);
\draw[thick] (Gm) -- (FR);

\end{tikzpicture}
\end{center}
Each circular node denotes an \(SU(k)\) gauge group, and the two rectangular boxes on the ends 
denote the left and right \(SU(k)\) flavor symmetries.  
There are \(N-1\) gauge nodes in total, connected by bifundamental hypermultiplets, while 
each gauge node also couples to a tensor multiplet whose scalar controls the gauge coupling 
along the tensor branch.  

Giving vevs to hypermultiplet scalars  
triggers Higgs-branch flows that break some of the flavor symmetry groups.  
These flows are characterized by the data of two nilpotent elements  
\(\mu_L,\mu_R \in \mathfrak{su}(k)\), associated respectively to the left and right flavor factors.  
Equivalently, they are labeled by two Young diagrams \(Y_L\) and \(Y_R\) with \(k\) boxes each.  
The vev pattern specified by \(\mu_L\) or \(Y_L\) determines how the left \(SU(k)\) flavor  
symmetry is broken, and similarly for the right one, see \autoref{fig:young-diagram} for an example.  Different choices of  
\((Y_L,Y_R)\) thus give rise to distinct interacting fixed points \(T_{N,k,\mu_L,\mu_R}\),  
each defining a separate conformal theory.

\begin{figure}
    \centering
    \begin{tikzpicture}[
  x=0.6cm,y=0.6cm,
  box/.style={draw,minimum width=0.55cm,minimum height=0.55cm},
  note/.style={draw,rounded corners,align=left,inner sep=6pt},
  lab/.style={font=\small}
]

\begin{scope}[shift={(0,0)}]
  
  \foreach \i in {0,1,2} \node[box] at (\i,0)   {};
  \foreach \i in {0,1,2} \node[box] at (\i,-1)  {};
  \foreach \i in {0,1}   \node[box] at (\i,-2)  {};
  \node[lab] at (1,-3.0) {$Y_L=[3,3,2]$};
\end{scope}

\node[anchor=west] (NL) at (5,-1.0)
  {
   $\mathrm{Comm}_{SU(k)}(\rho_{Y_L})
    = S\!\big(U(2)\times U(1)\big)
    \cong SU(2)\times U(1)$};

\draw[thick,->] (3.3,-1.0) -- (NL.west);

\end{tikzpicture}
    \caption{Example of a Young diagram $Y_L=[3,3,2]$ corresponding to a nilpotent element 
$\mu_L\in\mathfrak{su}(8)$. Each row of the diagram represents an irreducible 
$SU(2)$ representation of dimension equal to the row length.  
The multiplicity of equal row lengths determines the unbroken subgroup of $SU(k)$ 
commuting with the associated embedding $\rho_{Y_L}:SU(2)\hookrightarrow SU(k)$.  
In this example, there are two rows of length $3$ and one row of length $2$, giving the commutant
$\mathrm{Comm}_{SU(k)}(\rho_{Y_L}) = S(U(2)\times U(1)) \cong SU(2)\times U(1)$.  
Thus, the Young diagram encodes how the left flavor symmetry $SU(k)$ is broken 
along the Higgs branch of the corresponding six-dimensional $(1,0)$ SCFT.}
    \label{fig:young-diagram}
\end{figure}

The gravity duals of these Higgsed SCFTs are the aforementioned AdS$_7\times M_3$ solutions of massive IIA string theory \cite{Apruzzi:2013yva,Apruzzi:2015wna,Cremonesi:2015bld}. The $\mu_L$, $\mu_R$ encode the positions and charges of the internal D8/D6-branes, as explained in \cite{DeLuca:2018zbi}. In particular, the un-Higgsed case with $Y_L=Y_R=[1^k]$ corresponds to two D6-stacks with $F_0=0$, which in turn dualize to the $\mathbb{Z}_k$ singularities in M-theory. The fully-Higgsed case is $Y_L=Y_R=[k]$, and corresponds to two D8-branes with D6-charge $\pm k$. One can view the Higgsed cases as the result of a Myers-like effect, puffing up the D6s into D8-branes.
 
A seven-dimensional gauged supergravity description was proposed in \cite{DeLuca:2018zbi}. It contains the gravity multiplet together with two \(SU(k)\) vector  
multiplets, coming from the D6-stacks or from two fixed points of the orbifold $S^4/\mathbb Z_k$.  
The scalar sector sits in
\begin{align}\label{eq:Mscal}
\mathcal M_{\rm scal}=\frac{SO(3,2k^2)}{SO(3)\times SO(2k^2)}\,,\qquad
\Phi\in (\mathbf 3,\mathbf{Adj}_{SU(k)_L})\oplus(\mathbf 3,\mathbf{Adj}_{SU(k)_R}) .
\end{align}
The potential of this theory admits a discrete set of supersymmetric  
\(\mathrm{AdS}_7\) critical points.  
Each critical point is specified by the choice of two \(SU(2)\) embeddings into the two  
\(SU(k)\) gauge factors, precisely the data encoded by  
\((\mu_L,\mu_R)\) as
\begin{subequations}
\begin{align}
    \mathbf k &\to \mathbf k_1 + \cdots + \mathbf k_M,\\
    k&=\sum_{a=1}^M k_a,
\end{align}
\end{subequations}
where $\mu$ denotes a partition, $\mathbf k$ is the fundamental of $SU(k)$, and $\mathbf k_a$ are irreps of $SU(2)$ labeled by their dimension.

The scalars transform in the representation \((\mathbf 3,\mathbf{2k^2})\) of 
\(SO(3)\times SO(2k^2)\).  The \(SO(3)\) factor corresponds to the R-symmetry, 
while the \(SO(2k^2)\) acts on the adjoint indices of the two \(SU(k)\) gauge groups 
appearing in the truncated theory. Write the \(SO(3)_R\) triplet as \(\phi^i\) (\(i=1,2,3\)). The two gauge couplings are denoted
\(g_3\) for the \(SO(3)_R\) factor in the gauging and \(g_L,g_R\) for the two \(SU(k)\)’s; one can set one side to trivial when focusing on a single embedding.

Supersymmetric \( \mathrm{AdS}_7 \) vacua are captured by an \(SO(3)_R\)-invariant ansatz aligning
\(\phi^i\) along the chosen \(SU(2)\subset SU(k)\) embedding:
\begin{align}\label{eq:phi-vac}
\phi^i  =  \psi\, \sigma^i,\qquad [\sigma^i,\sigma^j]=\epsilon^{ijk}\sigma^k ,
\end{align}
where \(\psi\in\mathbb R\) is the single real amplitude and \(\{\sigma^i\}\) generate the chosen
\(\mathfrak{su}(2)\) subalgebra inside \(\mathfrak{su}(k)\). For a fixed \(Y\), define the quadratic Casimir parameter
\begin{align}\label{eq:kappa2}
\kappa^2  =  \sum_{a}\frac{k_a(k_a^2-1)}{12}\,,
\end{align}
which depends only on the partition. With this ansatz, it was shown \cite[Sec.~4.2]{DeLuca:2018zbi} that the scalar equations of motion reduce to a single algebraic relation for the invariant
\(\alpha=\psi\kappa\) as
\begin{equation}\label{eq:alpha-def}
\tanh(\alpha)=\frac{\kappa g_3}{g_L}\,.
\end{equation}
Comparing the cosmological constants in 7d and 10d, one can fix $(g_3/g_L)^2 = 12/(N^2 k)$.\footnote{This corrects an error in \cite[(4.16)]{DeLuca:2018zbi}.}
The vev is then fixed by the value of the Casimir \(\kappa\) of the embedded $SU(2)$.

The residual flavor symmetry is the commutant of the embedding,
\begin{align}\label{eq:HY}
H_Y=\mathrm{Comm}_{SU(k)}(\rho_Y)=S\!\Big(\prod_d U(f_d)\Big),
\end{align}
with \(S(\cdot)\) imposing overall determinant \(=1\) and $f_d$ denoting the number of rows of length $d$. Gauge bosons in
\(\mathfrak{su}(k)\ominus\mathfrak{h}_Y\) eat the corresponding Goldstone bosons.

\paragraph{Mass spectrum.}
Expanding around \(\phi^i=\psi\sigma^i\), the adjoint decomposes as
\begin{subequations}
\begin{align}\label{eq:dir-prod}
\mathrm{Adj}_{SU(k)}=(\mathbf k\otimes \overline{\mathbf k})\ominus 1
&\to \left(\Big(\bigoplus_{a=1}^{M}\mathbf k_a\otimes \mathbf k_a\Big)\ominus \mathbf 1\right)
\ \oplus\
2\!\!\!\bigoplus_{1\le b<a\le M}\!\!\mathbf k_a\otimes \mathbf k_b\\
&= \bigoplus_i \mathbf d_i.\label{eq:decomp-SUk}
\end{align}
\end{subequations}
Note that the adjoint is obtained by subtracting a trivial irrep $\mathbf 1$ from the square of the fundamental so has dimension $k^2-1$.

Each $SU(2)$ irrep $\mathbf {d}_i$ in \eqref{eq:decomp-SUk} produces two scalar towers with operator dimensions
\begin{subequations}\label{eq:spec12}
\begin{align}\label{eq:spec1}
\Delta=2d_i+4\qquad \text{ in }\mathbf{d}_i\mathbf{-2},\\\label{eq:spec2}
\Delta=2d_i+2\qquad \text{ in }\mathbf{d}_i\mathbf{+2},
\end{align}
\end{subequations}
where the first line is present only for $d_i>2$.
In addition, there are eaten Goldstone bosons at \(\Delta=6\) for \(d_i> 1\) with irrep $\mathbf d_i$. These follow from the mass matrix analysis around the BPS vacua in the 7d gauged supergravity and the standard holographic dictionary \cite[Table 1]{DeLuca:2018zbi}.  \eqref{eq:spec12} can be assigned \cite[Sec.~3.3]{Apruzzi:2019ecr} to a massive vector multiplet representation of $\mathfrak{osp}(6,2|1)$ for $d_i\neq1$, and to a massless vector multiplet for $d_i=1$ ($D_1$ in \cite[Sec.~4.7, 5.7.1]{Cordova:2016emh}).

Below we consider the special cases of $SU(2)$ embeddings:
\begin{itemize}\itemsep2pt
\item \emph{Trivial \(Y=[1^k]\):} \(\kappa=0\Rightarrow \psi=0\). This is the origin corresponding to the unbroken $SU(k)$ phase. The spectrum consists of $k^2-1$ tachyons with $\Delta=4$ at the BF bound \cite{Breitenlohner:1982bm, Breitenlohner:1982jf}.
\item \emph{Principal \(Y=[k]\):} $\kappa$ is maximized, corresponding to the fully Higgsed phase furthest from the origin. All scalars are massed up.
\end{itemize}

\paragraph{Naive count.}
For fixed \((N,k)\), the supersymmetric \(\mathrm{AdS}_7\) vacua are labeled by nilpotent data in the
left and right \(SU(k)\) flavor factors, i.e. by Young diagrams \(Y_L,Y_R\) with \(k\) boxes.
Naively, one would count the number of such vacua by counting partitions of \(k\), in the spirit of the previous naive count in \eqref{eq:V-naive}.

Let \(p(k)\) denote the number of partitions of the integer $k$. Counting the partitions on both left and right, we have $p(k)^2$ many vacua. By the Hardy--Ramanujan asymptotic,
\begin{align}
V_{N,k}(\hat\mu)\overset{\text{naive}}{=}p(k)^2\sim \frac{1}{48k^2}\exp\!\left(2\pi\sqrt{\frac{2k}{3}}\right),
\end{align}
so the naive count grows \emph{exponentially}:
\begin{equation}
\begin{split}
    \mathfrak{N}_{\mathrm{AdS}_7}^{(Q=8)} &\overset{\text{naive}}{\sim} \int_1^{\hat\mu^{-5/3}} dN\int_1^{\hat\mu^{-5/2}N^{-3/2}} dk  \frac 1 {k^2}\exp\!\left(2\pi\sqrt{\frac{2k}{3}}\right)\\
    &\sim \hat\mu^{35/6} \exp\!\left(2\pi\sqrt{\frac{2}{3}}\hat\mu^{-5/4}\right).
\end{split}
\end{equation}
In terms of the cosmological constant,
\begin{align}\label{eq:exp-Lambda-count}
    \sim |\hat\Lambda |^{35/12} \exp\left(2\pi \sqrt{\frac{2}{3}} |\hat\Lambda |^{-5/8}\right).
\end{align}

This exponential proliferation sharply contrasts with the power law scaling found in the \((N,k)\) discrete
count and motivates replacing the naive ``one per partition'' weight by the more meaningful volume-based measure on theory
space in the next subsection.

\paragraph{Total volume.}
We now consider the other extreme approach. Instead of counting the vacua individually, we consider the total volume of the scalar manifold for a fixed \((N,k)\). 

In general, scalar manifolds of supergravity must be quotiented by dualities when lifted to M-theory. These duality groups are arithmetic groups of the corresponding manifolds. In our case, we assume there is such a duality group such that the scalar manifold in M-theory becomes
\begin{align}
\widehat{\mathcal{M}}_{\mathrm{scal}} = \frac{O(3,\,2k^2;\mathbb R)}{O(3)\times O(2k^2)}\Big/ O(L,\mathbb Z),
\end{align}
where \(O(L,\mathbb Z)\) is the integral orthogonal group of a lattice \(L\) of signature \((3,2k^2)\). Note that the lattice choice matters: for unimodular \(L\) there is a canonical \(O(L,\mathbb Z)\), while for non-unimodular \(L\) the arithmetic group is not equivalent, hence the total volume depends on the choice of \(L\). However, the asymptotic behavior is similar for all choices.

For arithmetic quotients of orthogonal Grassmannians, the volume can be written as \cite[(4.6)]{Moore:2015bba}
\begin{align}
\mathrm{Vol}(\widehat{\mathcal{M}}_{\mathrm{scal}}) = \frac{\sigma(3+2k^2)}{\sigma(3)\,\sigma(2k^2)} m(L),
\end{align}
where \(m(L)\) is the mass of the genus of \(L\) obtained by Siegel mass formula \cite{Siegel, BELOLIPETSKY2005221} and $\sigma(p)$ is the volume of $O(p)$:
$$\sigma(p) = 2^{\frac{p+1}{2}} \prod_{j=1}^p \frac{(2\pi)^{\frac{j+1}{2}}}{\Gamma( \frac{j+1}{2})}.$$

We show in \autoref{app:Grassmannian} that the resulting total volume grows \emph{superexponentially} with \(k\),
\begin{align}
\mathrm{Vol}(\widehat{\mathcal{M}}_{\mathrm{scal}}) \sim k^{\,k^4},
\end{align}
far faster than any polynomial in \(k\).  
Thus, taking the entire scalar manifold volume $\mathrm{Vol}(\widehat{\mathcal{M}}_{\mathrm{scal}})$ as the weight $V_{N,k}(\hat\mu)$, the theory-space measure would be dominated by an uncontrolled superexponential factor.

Rather than using the full arithmetic quotient volume, it is more reasonable to restrict to a bounded region in the center containing all the vacua, in order to not count the parts of the scalar manifold that are not energetically accessible due to the cutoff $\hat\mu$. 

\paragraph{Volume of the large center ball.} At the origin of the scalar manifold $\mathcal M_{\text{scal}}$, there are $k^2-1$ tachyonic directions. This vacuum corresponds to $Y=[1^k]$. The goal is to estimate the effective volume of a finite-radius ball centered at the origin, which represents the physically accessible region of the space for a fixed $(N,k)$ configuration.

In particular, the ``center ball’’ is defined as the smallest ball in $\mathcal{M}_{\mathrm{scal}}$ that contains all vacua corresponding to partial Higgsings between the $SU(k)$ phase and the fully broken phase.

The geodesic distance between these phases follows from the $O(3,2k^2)$ invariant metric on the coset. 
The trajectory connecting the $SU(k)$ point to a vacuum point is a geodesic generated by an $O(3,2k^2)$ boost of rapidity $\alpha$ in a fixed timelike 3–plane. Choose an orthonormal set $u^i{}_r$, where $i$ runs over $SO(3)$ indices and $r$ over $SO(n)$, spanning the three negative directions that mix with the $SO(3)_R$ indices. Define the projector
\begin{align}
\Pi^{rs} = u^i{}^r u^i{}^s.
\end{align}
A convenient coset representative along this geodesic is then
\begin{equation}
L(\alpha) =
\begin{pmatrix}
\cosh\alpha\,\delta^{ij} & \sinh\alpha\,u^i{}_r \\
\sinh\alpha\,u^s{}_j &\ \ \  \delta^{rs} + (\cosh\alpha-1)\,\Pi^{rs}
\end{pmatrix},
\qquad L^{-1}(\alpha)=L(-\alpha).
\end{equation}

The sigma–model kinetic term is
\begin{equation}
\mathcal L_{\rm kin}
= -\frac12\,P_\mu^{ir} P^{\mu}{}_{ir},
\qquad
P_\mu^{ir} = (L^{-1}\partial_\mu L)^{ir}.
\end{equation}
Along the one–parameter trajectory $\alpha(x)$ one finds
\begin{equation}
P_\mu^{ir} = (\partial_\mu\alpha)\,u^{ir},\qquad
u^{ir}u_{ir}=3\,,
\end{equation}
so that
\begin{equation}
\mathcal L_{\rm kin}
= -\frac32\,(\partial_\mu\alpha)(\partial^\mu\alpha)\,.
\end{equation}
Thus the kinetic term induces a line element on field space along $\alpha$ as
\begin{equation}
ds^2 = 3\,d\alpha^2,
\end{equation}
and the geodesic distance between the $SU(k)$ point ($\alpha=0$) and a vacuum at rapidity $\alpha$ is
\begin{equation}
R = \int_0^\alpha \sqrt{3}\,d\alpha' = \sqrt{3}\,\alpha\,.
\end{equation}

Using the BPS relation
\begin{equation}
\tanh\alpha = \frac{\sqrt{12}\,\kappa}{N\sqrt{k}}\,,
\end{equation}
we finally obtain the radius from the origin as
\begin{equation}\label{eq:R-arctanh0}
R = \sqrt{3}\,\alpha
= \sqrt{3}\,\mathrm{arctanh}\!\left(\frac{\sqrt{12}\,\kappa}{N\sqrt{k}}\right)\,.
\end{equation} 

In the parametrization of the SCFTs, a priori it makes sense to take any value of $k$. However, one can see that the cases where the sum of the largest integers of the partitions $\mu_L$, $\mu_R$ is $>N$ are in fact redundant.  So we restrict our attention to cases where the largest integer in both partitions is $\le N/2$. In fact, for partitions that don't respect this restriction, the Casimir $\kappa_{L,R}$ are large, and as noted in \cite[Sec.~4.3]{DeLuca:2018zbi} the 7d theory appears to break down in that case: the cosmological constant does not reproduce the value expected from 10d. 

With this restriction, the partition that is furthest from the origin $[1^k]$ is $[k]$ when $k\le N/2$, and $[ k- (N/2) \lfloor 2k/N\rfloor, (N/2)^{\lfloor 2k/N\rfloor}]$ for $k>N/2$. The argument of the $\mathrm{arctanh}$ in \eqref{eq:R-arctanh0} is  $\kappa g_3/g_L= \sqrt{12}\kappa/N\sqrt{k}\sim k/N$ and $\sim 1/2$ in these two cases respectively. 
In both cases we can approximate
\begin{align}\label{eq:R-arctanh}
R \approx  \frac{6\kappa}{N\sqrt{k}} < \frac{\sqrt3}2 \,.
\end{align}

In Euclidean space $\mathbb R^d$, the volume of a fixed-radius ball decays superexponentially with dimension $d$ as
\begin{align}
    \mathrm{Vol}(B^{(d)}_R) = \frac{\pi^{d/2}}{\Gamma(d/2+1)} R^d \sim d^{-d/2}.
\end{align}
The same qualitative behavior persists for $O(3,2k^2)$ as $k\to\infty$.

In particular, we show in \autoref{app:Grassmannian} that the ball volume in $\frac{O(3,2k^2)}{O(3)\times O(2k^2)}$ is up to $O(1)$ constants
\begin{align}
    \mathrm{Vol}(B_R^{(3,2k^2)})\approx \frac{8\pi^{3k^2}}{\Gamma_3(k^2)}\times R^{6k^2}(2k^2)^{-3},
\end{align}
where $\Gamma_3(a)\equiv\pi^{3/2} \Gamma(a)\Gamma(a-1/2)\Gamma(a-1)$ is the multivariate gamma function. Here, the first factor denotes the angular surface area and the second factor is due to the radial direction. By using the Stirling asymptotic, we see that the volume decays superexponentially due to the shrinking angular surface area
\begin{align}
    \mathrm{Vol}(B_R^{(3,2k^2)}) \sim  k^{-6k^2}.
\end{align}

The physical moduli space volume must be divided by the gauge group volume associated with the $SU(k)$ symmetry of the scalar manifold. The volume is given for example in \cite[(2.5)]{Ooguri:2002gx}
\begin{align}\label{eq:vol-su}
    \mathrm{Vol}(SU(k))= \frac{\sqrt{k}(2\pi)^{\frac 1 2 k^2+\frac 1 2 k-1}}{G_2(k+1)}\sim k^{-\frac 1 2 k^2}
\end{align}
where $G_2$ is the Barnes $G$ function, with asymptotics \eqref{eq:G-as}. Therefore, the effective volume weight for fixed $(N,k)$ is
\begin{align}
\frac{\mathrm{Vol}(B_R^{(3,2k^2)})}{\mathrm{Vol}(SU(k))} \sim
k^{-\tfrac{11}{2}k^2}.
\end{align}
The result implies that as the number of dimensions $2k^2$ of the scalar manifold grows, the effective volume of the central ball shrinks extremely fast. 
Hence, even at small $k>1$, the contribution of each fixed $(N,k)$ sector to the total measure of vacua becomes superexponentially small. In other words, the count of vacua is dominated by only $k=1$ for each $N$. So we get a count that reproduces the count of the maximally supersymmetric case in \eqref{eq:count-maxSUSY}:
\begin{equation}
\begin{split}
    \mathfrak{N}_{\mathrm{AdS}_7}^{(Q=8)} (\hat\mu) &\sim \int_1^{\hat\mu^{-5/3}} dN\\
    &\sim \hat\mu^{-5/3}.
\end{split}        
\end{equation}

This result seems to be an important lesson we are learning: {\it Theories with large number of light or tachyonic modes contribute very little to the volume of all theories. } This is expected to be a general result because the diameter of the region in moduli for the validity of EFT where we compute the volume is expected to be bounded by order 1 in Planck units due to the distance conjecture.
\paragraph{Volume of small balls with changing dimensions.}
However, at each vacuum point, some of the $k^2$ tachyonic directions acquire mass. Since we already exclude massive directions, the relevant local theory space dimension equals the number of tachyonic directions $T$ at each vacuum, which changes at each point. We therefore compute the volume of the local $O(3,T(\mu))$ ball of radius $R(\mu)$ centered at each vacuum, where $R(\mu)$ is set by the distance to the nearest neighboring vacuum.

Recall that each vacuum is labeled by a partition $\mu$ of $k$,
\begin{align}
\mathbf k \to \mathbf k_1+\mathbf k_2+\dots+\mathbf k_M,
\end{align}
and the spectrum is given by \eqref{eq:spec1} and \eqref{eq:spec2} for each $\mathbf{d}_i$ that shows up in the decomposition of the adjoint \eqref{eq:decomp-SUk}.

Tachyons arise in $\mathbf 3$ representations of $SU(2)$, corresponding to each instance of $d=1$ in \eqref{eq:decomp-SUk}. From the direct product structure of \eqref{eq:dir-prod}, self-tensor terms of the form $\mathbf{k}_a\otimes\mathbf{k}_a$ yield $3(M-1)$ tachyons, and cross terms of the form $\mathbf{k}_a\otimes\mathbf{k}_b$ yield tachyons only if $\mathbf{k}_a=\mathbf{k}_b$.
The number of $\mathbf{k}_a$ with dimension $d$ is the number of fundamental flavors $f_a$ in the quiver. The total number of tachyons can then be written as  $3\sum f_a^2-3$, or in other words 
\begin{equation}\label{eq:Tmu}
    T(\mu) = 3 \,{\rm dim} H_Y\,,
\end{equation}
the dimension of the unbroken flavor group \eqref{eq:HY}. Indeed, recalling our remark below \eqref{eq:spec12}, the $d_i=1$ scalars sit in massless vector multiplets. The holographic duals to these tachyons are the so-called \emph{(hyper-)momentum map} operators associated to a flavor symmetry.

For a given partition $\mu=\{k_i\}$, recall the definition of $\kappa^2$ from \eqref{eq:kappa2}.
Neighboring vacua are obtained by transferring one unit between two parts, 
$k_i\to k_i-1$, $k_j\to k_j+1$. The resulting difference is 
\begin{align}
\kappa_{\mu'}^2-\kappa_\mu^2 = \frac{k_j(k_j+1)-k_i(k_i-1)}{4}.
\end{align}
The largest change occurs when $k_i=k_{\min}$ and $k_j=k_{\max}$
\begin{align}
\sup_{i,j}(\kappa_{\mu'}^2-\kappa_\mu^2)=\frac{k_{\max}(k_{\max}+1)-k_{\min}(k_{\min}-1)}{4}.
\end{align}
This is related to the difference between the radii of the two vacua from the origin. 

Thus the maximum geodesic radius to a neighboring vacuum is
\begin{align}
\delta R \approx 6\Big[\mathrm{arctanh}\!\Big(\frac{\kappa_{\mu'}}{N\sqrt{k}}\Big)
- \mathrm{arctanh}\!\Big(\frac{\kappa_{\mu}}{N\sqrt{k}}\Big)\Big].
\end{align}
For large $N$, the arguments are small, so
\begin{equation}\label{eq:R-approx}
\begin{split}
\delta R &\approx 6\frac{\kappa_{\mu'}-\kappa_\mu}{N\sqrt{k}}
= 6\frac{\kappa_{\mu'}^2-\kappa_\mu^2}{N\sqrt{k}(\kappa_{\mu'}+\kappa_\mu)} \nonumber\\
&= \frac32\frac{k_{\max}(k_{\max}+1)-k_{\min}(k_{\min}-1)}{N\sqrt{k}(\kappa_{\mu'}+\kappa_\mu)}.
\end{split}    
\end{equation}

To understand whether ignoring the angular separations between neighboring vacua as we have done above is justified, we work directly with the coset geometry. A point is parametrized by a boost parameter $\alpha$ and a unit embedding direction $u^{i}{}_{r}$. If we vary only the direction $u$ at fixed $\alpha$ (i.e. $d\alpha=0$ and $du\neq 0$), the components of $P$ are
\begin{align}
P^{ir}= \sinh\alpha\, du^{ir}\approx \alpha d u^{ir} = \frac{R}{\sqrt{3}}\frac{d\sigma^i{}_r}{|\kappa|},
\end{align}
where $\alpha \ll 1$. This plays the same role as $R d\theta$ in flat polar coordinates. With $R=\sqrt{3}\alpha$, the angular contribution to the line element becomes
\begin{align}
R^{2}\delta\theta^{2}
\approx \frac{R^2}{3}\frac{\mathrm{Tr}\delta \sigma^i \delta \sigma^i}{\kappa^2}\sim \frac{k^2_{\max}}{N^2k}.
\end{align}
To estimate $\mathrm{Tr}\delta\sigma^{2}$, note that for each vacuum the $SU(2)$ embedding is specified by a block decomposition of the $k\times k$ fundamental representation into irreducible spins. Changing the partition by one unit (moving a single box between neighboring rows) shifts the block dimension from $k_i$ to $k_i\pm 1$, which corresponds to replacing a spin-$j$ representation by spin-$j\pm\tfrac12$. The associated SU(2) generators $\sigma^{i}$ therefore change only inside that block. Inside a spin-$j$ block, the matrix elements are of size $O(\sqrt{j})\sim O(\sqrt{k_i})$, coming from the top and bottom ladder elements. Consequently the squared norm for changing $k_{\max}$ to $k_{\max} +1$ scales as
\begin{align}
{\rm Tr}(\delta\sigma^{2})
\sim k_{\max}^{2}.
\end{align}

For $\delta R$ we have
\begin{align}
    \delta R \sim \frac{\delta \kappa}{N \sqrt{k}} \sim \frac{k_{\max}}{N \sqrt{k}},
\end{align}
so we get the angular line element
\begin{align}
\frac{R^{2}\delta\theta^{2}}{\delta R^{2}}
\sim \frac{1}{k_{\max}^{2}}.
\end{align}
For points far away from the origin, we have $k_{\max}\sim k$, so this implies the radial difference is a good approximation for the radius of the ball.

If instead $k_{\max}=O(1)$, then $\kappa\sim\sqrt{k}$ and the vacua lie near the origin with $R\sim 1/N$. In that case, the angular distance potentially matters, but we can instead estimate the ball radius as the distance from the origin $\delta R\approx R\sim 1/N$. In either regime the characteristic spacing is $1/N$.

Using \autoref{app:Grassmannian}, the corresponding ball volume for $T$ tachyonic directions is given by
\begin{align}
    \mathrm{Vol}(B_{\delta R}^{(3,T)})\approx \frac{8\pi^{3T/2}}{\Gamma_3(T/2)} \delta R^{3T}T^{-3}.
\end{align}
Now we consider the residual gauge groups. Each vacuum has an unbroken gauge subgroup $H_Y$ given by \eqref{eq:HY}. To obtain the physical moduli-space volume, divide by the gauge-group volume:
\begin{align}
\frac{\mathrm{Vol}(B_{\delta R}^{(3,T)})}{\mathrm{Vol}(H_Y)}.
\end{align}

Summing over all partitions $\mu\vdash k$, the total weight is
\begin{align}
V_{N,k}=\sum_{\mu\vdash k}W(\mu),\qquad
W(\mu)=\frac{\mathrm{Vol}(B_{\delta R(\mu)}^{(3,T(\mu))})}{\mathrm{Vol}(H_{Y(\mu)})},
\end{align}
We group terms by partition length $\ell(\mu)=M$: 
\begin{align}
\sum_{\mu\vdash k}W(\mu)=\sum_{M=1}^k\mathfrak W(M),\\
\mathfrak W(M)\equiv\sum_{\ell(\mu)=M}W(\mu).
\end{align}

Let $SU(f)$ denote the largest unbroken factor in $H_Y$. The approximate form of the weight is then
\begin{align}
W(\mu)\approx
\frac{G_2(f+1)}{\sqrt{f}(2\pi)^{\frac12f^2+\frac12f-1}}
\times
\frac{\pi^{3T/2}}{\Gamma_3(T/2)}\frac{\delta R^{3T}}{T^3}.
\end{align}
The Barnes G-function has asymptotics \eqref{eq:G-as}, so
$G_2(f+1) \sim f^{f^2/2}$; since $T>f$,\footnote{Except for the case $\mu=[k]$, when $T=0$.} the dominating factor is $\Gamma_3(T/2)$, which makes the weight decay superexponentially fast in $T$. This means that $\mathfrak W(M)$ decays sharply with $M$. In fact, the numerical study over the combinatorial sum in the supplementary Mathematica notebook reveals that the decay starts immediately, and only the $M=1$ partition contributes significantly, which is simply
\begin{align}
    V_{N,k}=\sum_{\mu\vdash k} W(\mu) \sim 1.
\end{align}
We now see that the naive weight assignment \eqref{eq:V-naive} turned out to be the correct weight after all. We claim that this is a generic behavior: as the number of degrees of freedom and hence the dimension increases, the volume of the spheres decrease, so only the vacuum with the least number of light scalars contributes.  From now on, we will assume that this phenomenon is generic and the relevant directions do not change the leading volume factor and thus use the discrete count weight
\begin{align}\label{eq:weight-1}
    \boxed{V_{N,k,\dots}\sim 1.}
\end{align}

The corresponding $\mathrm{AdS}_7$ count is therefore \eqref{eq:count-AdS7-lowSUSY}
\begin{align}
    \mathfrak{N}_{\mathrm{AdS}_7}^{(Q=8)}(\hat\mu) \sim \hat\mu^{-5/2}.
\end{align}

\paragraph{Correction to the weight. } To see the magnitude of the correction from the other terms in the sum, consider the $M=2$ terms. We have $T=3$ with $k=k_1+k_2$.  
Then $\kappa\sim k^{3/2}$ and
\begin{align}
\delta R\sim \frac{k_i^2}{N\sqrt{k}\kappa}\sim \frac 1 N.
\end{align}
Therefore
\begin{align}
W(\mu)\big|_{M=2}&\sim \delta R^9\sim \frac 1 {N^9},\\
\mathfrak W(2)&\sim kW(\mu)\sim kN^{-9},
\end{align}
since there are $\lfloor k/2 \rfloor\sim k$ partitions of length $\ell(\mu)=2$. So the first correction to the counting weight of $(N,k)$ is
\begin{align}
    V_{N,k} \sim \sum_{\ell(\mu)=1,2} W(\mu) \sim 1+kN^{-9}.
\end{align}

\subsection{\texorpdfstring{$\mathrm{AdS}_5$}{AdS5}}\label{sec:AdS5}

\subsubsection{Sphere quotients}\label{sec:AdS5-sphere}

We consider type IIB string theory on a supersymmetric sphere quotient\footnote{We consider abelian orbifolds here.  One can also consider non-abelian ones, however we do not expect the general conclusion to change dramatically for the general case. Indeed the largeness of infinite series of non-abelian ones comes from abelian subgroups of it.}
\begin{align}
    \mathrm{AdS}_5 \times S^5/(\mathbb Z_{k_1}\times \mathbb Z_{k_2}),
\end{align}
with $N$ units of $F_5$ on the $S^5/(\mathbb Z_{k_1}\times \mathbb Z_{k_2})$. The gravitini transform as $\mathbf{4}\oplus \overline{\mathbf{4}}$ under the spin-lift of the sphere isometries $Spin(6)\cong SU(4)$, so we preserve supersymmetry by choosing
\begin{align}
    \mathbb Z_{k_1} \times \mathbb Z_{k_2} \subset SU(3) \subset SU(4).
\end{align}
Assuming that the weight associated to each $(N,k_1,k_2)$ is $V_{N,k_1,k_2}\sim 1$ as per \eqref{eq:weight-1}, it is enough to do a discrete count. 

By the fundamental theorem of finite abelian groups, any finite abelian group is canonically written as a product of cyclic groups whose orders divide each other, by moving factors from one side to the other. To avoid overcounting, we always assume that
\begin{align}
    k_2 \mid k_1.
\end{align}
\paragraph{Scales.}
Let $R$ be the radius of $S^{5}$. Flux quantization and dimensional reduction give
\begin{align}
M_{10}^{4}\int_{S^{5}/(\mathbb Z_{k_{1}}\times\mathbb Z_{k_{2}})}F_{5}\sim M_{10}^{4}\frac{R^{4}}{k_{1}k_{2}}\sim N,\qquad
M_{5}^{3}\sim M_{10}^{8}\frac{R^{5}}{k_{1}k_{2}}.
\end{align}
Hence the useful dimensionless relations
\begin{align}
M_{10}R\sim (Nk_{1}k_{2})^{1/4},\qquad
(M_{5}R)^{3}\sim N^{2}k_{1}k_{2},\qquad
\frac{M_{5}}{M_{10}}\sim N^{5/12}(k_{1}k_{2})^{1/12}. \label{eq:AdS5-scales}
\end{align}
\paragraph{Towers.}
The $S^{5}$ KK tower sets
\begin{align}
\frac{m_{\rm KK}}{M_{5}}\sim\frac{1}{M_{5}R}\sim N^{-2/3}(k_{1}k_{2})^{-1/3}>\hat\mu. \label{eq:AdS5-KK}
\end{align}
We must also check wrapped $p$-branes on $q$-cycles of $S^{5}/(\mathbb Z_{k_{1}}\times\mathbb Z_{k_{2}})$. For $q\ge 2$, the minimal $q$-cycle volume scales as $R^{q}/(k_{1}k_{2})$, therefore
\begin{align}
    T_{p-q} \sim M_{10}^{p+1} \frac{R^q}{k_1k_2}.
\end{align}
In dimensionless terms, the associated spacetime $(p-q)$-brane tower scale is
\begin{align}
    \frac{T_{p-q}^{\frac 1 {p-q+1}}}{M_{5}} \sim \frac{M_{10}}{M_5} \left(\frac{(M_{10}R)^q}{k_1k_2}\right)^{\frac 1 {p-q+1}}\sim N^{-\frac{5}{12} + \frac{q}{4(p-q+1)}} (k_1k_2)^{-\frac{1}{12} + \frac{q-4}{4(p-q+1)}} \gtrsim \hat\mu.
\end{align}
Comparing with the exponents in \eqref{eq:AdS5-KK}, we see that the exponent of $N$ for the KK tower is always smaller. Also, explicitly comparing the exponents of the $k_1k_2$ term for the branes of IIB $p=3,5,7$ over cycles with $2\leq q \leq \min(p,5)$, we find that they are no lighter than the KK tower.

For $q=1$, the smallest $1$-cycle is $R/k_1$, since the count is over $k_1\geq k_2$. Then
\begin{align}
    \frac{T_{p-1}^{\frac{1}{p}}}{M_5} \sim \frac{M_{10}}{M_5} \left(\frac{M_{10}R}{k_1}\right)^{\frac{1}{p}} \sim N^{-\frac{5}{12}+\frac{1}{4p}}k_1^{-\frac 1 {12}+\frac 1 {4p}}k_2^{-\frac{1}{12}-\frac{3}{4p}}\gtrsim\hat\mu.\label{eq:AdS5-F1}
\end{align}
It turns out that this inequality is not implied by the KK tower inequality. In fact, we show in \autoref{app:Motzkin} that the inequalities \eqref{eq:AdS5-F1} for $p=1$ and \eqref{eq:AdS5-KK} form the minimal set of inequalities that imply all other inequalities through an application of Farkas lemma. These are the tower scales from KK and the F1/D1 strings wrapping a shortened equator of the orbifold.

\paragraph{Orbifold choices and multiplicities.}
We will show that given $k$ for $\mathbb Z_{k}$, there are $\psi(k)$ many $\mathbb Z_{k}\subset (\mathbb Z_{k})^2\subset U(1)^2$ choices, where $\psi$ is the Dedekind psi function.

First, count the number of order-$k$ elements $(x,y)$ in $(\mathbb Z_{k})^2$. If a prime $p|k$, then an order-$k$ element is a pair $(x,y)$ where at least one of the entries is not divisible by $p$. The ratio of such pairs without a $p$ divisor to all pairs is $1-1/p^2$. Doing this for each $p|k$, we get
$$k^2 \prod_{p|k} \left( 1-\frac{1}{p^2} \right). $$
Second, we need to determine which order-$k$ elements generate the same $\mathbb Z_{k}$ group. To do this, we divide the above count by the number of units in $\mathbb Z_{k}$, which is given by the Euler totient function
$$|\mathbb Z_{k}^\times|=\varphi(k) = k \prod_{p|k}\left( 1-\frac{1}{p} \right).$$
Therefore
$$\#\text{ of }\mathbb Z_{k} = \frac{k^2 \prod_{p|k} \left( 1- \frac{1}{p^2} \right)}{k \prod_{p|k}\left( 1-\frac{1}{p} \right)}=k \prod_{p|k} \left( 1+\frac{1}{p} \right)=\psi(k).$$
It is known that for large $k$, the Dedekind psi function asymptotes as \cite[Thm.~6.4]{mccarthy}
\begin{equation}\label{eq:psi-as}
    \psi(k) \sim k\,.
\end{equation}
We therefore conclude that there are $\sim k_1k_2$ many choices of $\mathbb Z_{k_1}\times \mathbb Z_{k_2}\subset U(1)^2$ in the maximal torus of $SU(3)$.

Lastly, we need a density measure for the divisors $k_2< k_1$ such that $k_2\mid k_1$. Let $\tau(k)$ denote the divisor function, which counts the number of divisors of $k$. Its mean value is asymptotically logarithmic \cite[Thm.~2.3]{montgomery-vaughan}
\begin{align}
    \tau(k) \sim \log k.
\end{align}
Therefore the associated measure is
\begin{align}
    d\tau \sim \frac{dk}{k}
\end{align}
such that
\begin{align}
    \int_1^k d\tau \sim\tau(k).
\end{align}

\paragraph{Cylindrical decomposition.}
The integration region is defined by the bound on the KK tower scale \eqref{eq:AdS5-KK}
\begin{align}\label{eq:AdS5-ine1}
    N^{-2/3}(k_{1}k_{2})^{-1/3}\gtrsim\hat\mu,
\end{align} 
and the F1 tower wrapping the equator $1$-cycle \eqref{eq:AdS5-F1}
\begin{align}\label{eq:AdS5-ine2}
    N^{-\frac 1 6}k_1^{\frac 1 6}k_2^{-\frac 5 6}\gtrsim\hat\mu,
\end{align}
in addition to the positivity of the discrete data
\begin{align} \label{eq:AdS5-ine3}
    N,k_1,k_2 \geq 1,
\end{align}
and the divisor condition for $k_2$
\begin{align}\label{eq:AdS5-ine4}
    k_1 \gtrsim k_2.
\end{align}
All additional wrapped-brane bounds are implied by the bounds above. 

\begin{figure}
    \centering
    \includegraphics[width=0.4\linewidth]{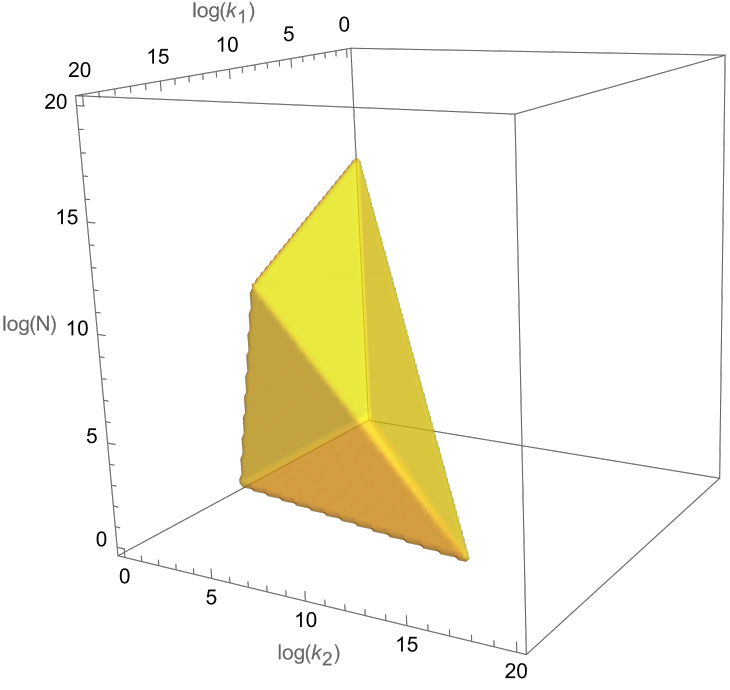}
    \caption{The integration region in $(\log(k_1),\log(k_2),\log(N))$ variables for the number of theories with tower scales above the cutoff $\hat\mu$ obtained from \eqref{eq:AdS5-ine1}, \eqref{eq:AdS5-ine2}, \eqref{eq:AdS5-ine3}, \eqref{eq:AdS5-ine4}. For the plot we chose $\log\hat\mu =-10$.}
    \label{fig:ads5-region}
\end{figure}

To be able to integrate this region, we reduce the above system of inequalities to a triangular set of inequalities by cylindrical decomposition; this reduction is carried out in Mathematica and the explicit form is too long to display. The resulting region is shown in \autoref{fig:ads5-region}.
\paragraph{AdS count.}
The multiplicities from orbifold choices give an integrand weight $k_{1}k_{2}$ in the continuum approximation. The count is therefore
\begin{align}
\mathfrak N_{\mathrm{AdS}_{5}}^{(Q=4)}(\hat\mu)
\sim\int dNdk_1\frac{dk_{2}}{k_{2}}  k_{1}k_{2}.
\end{align}
Evaluating, we get the count
\begin{align}\label{eq:AdS5-count-Q4}
\mathfrak N_{\mathrm{AdS}_{5}}^{(Q=4)}(\hat\mu)\sim \hat\mu^{-9/2}.
\end{align}
In terms of the cosmological constant,
\begin{align}
    \sim\hat|\Lambda|^{-9/4}.
\end{align}
Using $(M_{5}R)^{3}\sim c$ from \eqref{eq:AdS5-scales} and $M_{5}R\sim \hat\mu^{-1}$ at the boundary, this maps to the CFT count with central charge less than $\hat c$ as
\begin{align}
\mathfrak N_{\mathrm{CFT}_{4}}^{(Q=4)}(\hat c)\sim \hat c^{3/2}.
\end{align}
\paragraph{Special case ($Q=8$).}
Letting $k_2=1$ reduces to the special case with the choice $\mathbb Z_{k_{1}}\subset U(1)\subset SU(2)$. In this case, there is a unique embedding of $\mathbb Z_{k_1}$ for each $k_{1}$. The measure becomes $dNdk_{1}$, and we get the counts
\begin{align}
\mathfrak N_{\mathrm{AdS}_{5}}^{(Q=8)}(\hat\mu)\sim \hat\mu^{-2},\\
\qquad
\mathfrak N_{\mathrm{CFT}_{4}}^{(Q=8)}(\hat c)\sim \hat c^{2/3}.
\end{align}
Again, in terms of the cosmological constant we have
\begin{align}
    \sim |\hat\Lambda |^{-1}.
\end{align}

\subsubsection{Class S}
\label{sub:s}

The so-called \emph{class S} of ${\mathcal N}=2$ SCFTs is obtained by compactifying the ${\mathcal N}=(2,0)$ theory on a Riemann surface $\Sigma$ of genus $g$, possibly with $n$ punctures. 
The pants decomposition of $\Sigma$ is obtained by gluing $-\chi=2g-2+n$ three-punctured spheres with tubes. In the field theory each sphere is associated to a \emph{trinion} $T_N$ theory with $SU(N)^3$ flavor symmetry, a tube to the circle reduction of the 5d $SU(N)$ gauge theory, and their gluing to a gauging of the flavor symmetries. Each $SU(N)$ can be Higgsed as dictated by a partition of $N$, reminiscent of the AdS$_7$ case.

On the AdS side, the case without punctures is dual to the Maldacena--Nu\~nez solution \cite{Maldacena:2000mw}, whose internal space consists of a (topological) $S^4$ fibered over $\Sigma$. Punctures correspond to the addition of M5-brane stacks. Higgsing corresponds to them changing positions along the $S^4$, and possibly acquiring a NUT charge. The whole pattern is very reminiscent of the AdS$_7$ case (and of other cases, as we will comment below in \autoref{sub:disc}). 
These solutions with punctures are known locally around each puncture, or on $\Sigma=S^2$ with many simple punctures \cite{Gaiotto:2009gz}. 

Unfortunately a 5d theory similar to the 7d one in \cite{DeLuca:2018zbi} is not known. While a remarkable consistent truncation has been found recently \cite{Bhattacharya:2024tjw}, here we would ideally need a theory with a gauge group $SU(N)^n$, realizing all the vacua corresponding to all ways to Higgs the punctures. 

\paragraph{Scales.}
We take the naive AdS radius $L\sim l_p$; the warping is then $e^A \sim N^{1/3}$, where $N= \frac1{(2\pi)^3}\int_{S^4}G$. The internal metric scales as $d s^2_6 \sim N^{2/3}l_p^2 (ds^2_\Sigma + ds^2_4)$. We then have 
\begin{equation}
	M_5^3 = l_p^{-9} \int_{M_6} e^{3A} \sqrt{g_6}\sim l_p^{-3} N^3 |\chi| 
\end{equation}
where $\chi = 2-2g-n$ actually needs to be negative. From this,
\begin{equation}
	M_5 l_p = N |\chi|^{1/3}\,.
\end{equation}

\paragraph{Towers.}
The KK spectrum is expected to be of order $1/L$. Demanding that such modes be heavier than the cutoff $\mu$, we obtain 
\begin{equation}
		M_5 l_p = N^3|\chi| < \hat \mu^{-1} = \hat c^{1/5}\,. 
\end{equation}
There can be much smaller eigenvalues for particular choices of $\Sigma$. The \emph{systole} $\delta$ is the length of the smallest closed geodesic (in the dimensionless metric $ds^2_\Sigma$). When $\delta$ is small and \emph{separating}, namely $\Sigma$ minus the geodesic disconnects in two disjoint open surfaces, a small KK mass appears. A possible estimate leads to\footnote{The spin-two part of the spectrum was analyzed in \cite{DeLuca:2021ojx}. Th.~4.1, 4.2 in that reference give $h_1/4 \le m_1\le 4\max\{h_1,\sqrt{h_1/L} \}$; the \emph{Cheeger constant} $h_1$ is the smallest value of perimeter/area over any open surface $B\subset \Sigma$ such that $\text{area}(B)<\frac12(\text{area}\Sigma)$. With a small separating systole, if one of the two halves is small, we can estimate $h_1 \sim \delta/l_p$.
The universal part (present for any $\Sigma$) can be found in \cite{Bhattacharya:2024tjw}.} 
\begin{equation}\label{eq:delta-KK}
	\delta > N^2|\chi|^{2/3} \hat \mu^2\,.
\end{equation}
This actually does not have much effect on the final count.

We also impose that the brane tensions be larger than the cutoff scales. It turns out that they are all redundant except for
\begin{equation}\label{eq:delta-M22}
	\delta > |\chi|^{1/3}\hat\mu\,,
\end{equation}
which comes from M2-branes along the smallest closed geodesic in $\Sigma$, times a $S^1\times S^4$.

\paragraph{Moduli.}

The moduli space of these solutions is simply that of Riemann surfaces,  ${\mathcal M}_{g,n}$. Imposing that the reduced action be of the form
\begin{equation}\label{eq:S5}
    S_5 \supset m_5^3(\int d^5 x \sqrt{g}R -\frac1\kappa (\partial \phi)^2)
\end{equation}
for some factor $\kappa$ and using 
\cite[(3.12)]{Tachikawa:2017aux} we find 
\begin{equation}\label{eq:met-Mgn}
	g_{I\bar J}= \frac \kappa{\pi|\chi|} g^\mathrm{WP}_{I\bar J}\,
\end{equation}  
with $g^\mathrm{WP}_{I\bar J}$ the Weil--Petersson metric. 

The WP volume of ${\mathcal M}_{g,n}$ is given in \cite{mirzakhani-zograf} as $\mathrm{Vol}({\mathcal M}^\mathrm{WP}_{g,n})\sim C g^{-1/2}|\chi-1|! (4\pi^2)^{|\chi-1|}$ for a constant $C$. Thus the volume according to \eqref{eq:met-Mgn} is given by
\begin{align}\label{eq:volMgn}
	\nonumber\log\mathrm{Vol}({\mathcal M}_{g,n}) &\sim  |\chi|(\log|\chi|-1)+|\chi|\log(4\pi^2)-(3g+n)(\log|\chi|+\log(\pi/\kappa)) \\
	&\sim -g \log(2g+n)+ \alpha g+ (\beta-\log\kappa) n)
\end{align}
where $\alpha=\log(16\pi)\sim 1.92$, $\beta= \log(4\pi)-1\sim 1.53$. 

In the case with no punctures, the $-g\log(g)$ quickly overpowers the $\alpha g$ term, and the volume decreases fast. When punctures are present, the leading term in $n$ is $(\beta - \log\kappa)n$. So the volume decreases if $\kappa >e^\beta \sim 4.62$. 

In the case without punctures, the part of moduli space where the systole $\delta$ is larger than $\epsilon$ is also known \cite{mirzakhani-petri}: 
\begin{equation}\label{eq:Mdelta}
	\frac{\mathrm{Vol}_{\delta > \epsilon}({\mathcal M}_g)}{\mathrm{Vol}({\mathcal M}_g)} = e^{-\lambda(\epsilon)} \, ,\qquad \lambda(\epsilon)= \int_0^\epsilon (\sinh(t)-1)dt\,.
\end{equation}

\paragraph{Tachyons.}

We cannot count the number of tachyons directly in gravity as directly as we did in AdS$_7$, because of the already noted lack of an AdS$_5$ analogue of the supergravity model of \cite{DeLuca:2018zbi}. We will proceed using some knowledge of the dual SCFTs, and of superconformal multiplets.

For every $k=2,\,\ldots,\,N$, there are Coulomb branch operators with scaling dimension $\Delta=k$ (and $r=2\Delta$). These are relevant for $k=2,3$, and thus give rise to tachyons (above the BF bound). When the punctures are all maximal (associated to partition $[1^N]$), each trinion contributes $k-2$ of them; each tube, one (which is just ${\rm Tr}(\Phi)^k$). So in total there are $(k-2)(2g-2+n)+3g-3+n= (2k-1)(g-1)+(k-1)n$. When a puncture is Higgsed, its contribution drops \cite[Fact 5.14]{Tachikawa:2013kta}, for example to zero for a minimal puncture (partition $[N-1,1]$). 

The total  Higgs branch dimension is $4(N-1)+ 2nN(N-1)$, with $n$ maximal punctures. When punctures are Higgsed, this number drops by the dimension of the orbit associated to the associated partition \cite[Fact 5.12]{Tachikawa:2013kta}; this is similar to six dimensions \cite{Mekareeya:2016yal}.
This does not give us directly the desired number of tachyons, because not all of the Higgs branch generators are relevant. Moreover, in the Higgs branch there can be relations among operators, imposed by nilpotency of the vev. 

Fortunately, we can proceed by adapting the AdS$_7$ results reviewed earlier. Higgsing a puncture with a nilpotent vev identifies an $SU(2)\subset SU(N)$, whose diagonal with $SU(2)_R$ emerges as the R-symmetry group for the new SCFT. This predicts that each block of dimension $d$ generates representations $\mathbf{d}\mathbf{+2}$, $\mathbf{d}\mathbf{+2}$, $\mathbf{d}$ as in \eqref{eq:spec12}.

Now we can recall the structure of supersymmetry multiplets. The relevant 5d supergravity would be the gauged half-maximal one, with $\mathcal{N}=4$ supersymmetry: there are two symplectic Majorana spinor pairs as supersymmetry parameters.  A model with the gravity multiplet and $m$ vector multiplets has $1+5m$ scalars, parameterizing 
\begin{equation}
    \mathcal{M}_{\rm scal} = SO(1,1)\times \frac{SO(5,m)}{SO(5)\times SO(m)}\,.
\end{equation}
A supergravity similar to the 7d one in \cite{DeLuca:2018zbi} would be gauged so as to promote the $m$ vectors into a ${\rm dim}SU(N)^n$ gauge group. 

Around each AdS$_5$ vacuum, the spectrum organizes itself in superconformal multiplets. A vector multiplet (known as $B_1\bar B_1$ in \cite[Sec.~4.6,\,5.5.2]{Cordova:2016emh}) contains a vector in the $\mathbf{d}$ with $\Delta=d+2$, and scalars:
\begin{subequations}\label{eq:vm5}
\begin{align}
\label{eq:vm5-1}
&\Delta=d+3\quad \text{ in }\mathbf{d}\mathbf{-2},\text{ with }r=0\,,\\
\label{eq:vm5-2}
&\Delta=d+1\quad \text{ in }\mathbf{d}\mathbf{+2},\text{ with }r=0\,;\\
&\Delta=d+2\quad \text{ in }\mathbf{d},\text{ with }r=+2\,,\\
&\Delta=d+2\quad \text{ in }\mathbf{d},\text{ with }r=-2\,,
\end{align}
\end{subequations}
where $r$ is the $U(1)_R$ charge. So there is a total of $(d+2+(d-2)+d+d=4d$ scalars; in other words, four per vector; one scalar has been eaten in a Higgs mechanism.

The case $d=1$ is the massless vector multiplet; \eqref{eq:vm5-1} is missing, and there is a total of $3+1+1=5$ scalars per vector, with $\Delta=2,3,3$ respectively; so they are all relevant. The triplet with $\Delta=2$ is the moment map, as in the AdS$_7$ case above. The $\Delta=3$ have $r\neq0$, and the RG flow generated by their vev would break supersymmetry.

Around a general Higgsed vacuum, there are $G_{\rm unbroken}$ massless vector fields,
and $\dim SU(N)^n-\dim G_{\rm unbroken}$ massive vector multiplets. For $d=2$, the latter give additional relevant operators with $\Delta=3$ from \eqref{eq:vm5-2}. 

For example, a maximal puncture (partition $[1^N]$) contributes $5(N^2-1)$ relevant operators with $\Delta=2$; a closed puncture $[N]$ contributes no relevant operators. 

\paragraph{Count.}

Let us first consider the case without punctures, $n=0$. Let us also ignore the requirement \eqref{eq:delta-KK} for now. Using \eqref{eq:volMgn}:
\begin{equation}
	\sum_{g,N <(\hat c/|\chi|)^{1/3}} {\rm Vol}(\mathcal{M}_{g})\sim 
    \sum_{g,N <(\hat c/|\chi|)^{1/3}} e^{g(\alpha - \log g) - \lambda(\delta_0)} \, ,\qquad \delta_0 = (|\chi|/\hat c)^{1/3}\,.
\end{equation}
As commented earlier, $e^{g(\alpha - \log g)}$ decays quickly, and is tiny already for $g= g_0 \sim 10$. So the sum becomes from $g=2$ to $g_0$. Now take $\hat c\gg 2g_0$; $\delta_0$ is small, and \eqref{eq:Mdelta} tells us $\lambda(\delta_0)\sim 1$: there is no systole suppression. The sum over $N$ reduces to $\sum_{g=2}^{g_0} (\hat c/|\chi|)^{1/3}e^{g(\alpha - \log g)}$, so
\begin{equation}\label{eq:g-count}
	\sum_{g,N <(\hat c/|\chi|)^{1/3}} {\rm Vol}(\mathcal{M}_{g})\sim \hat c^{1/3}\,.
\end{equation}
If we do take into account \eqref{eq:delta-KK}, we need to divide the sum over $N$ into the range over $[1,(\hat c/2g)^{1/6}]$ and over $[1,(\hat c/2g)^{1/6}]$; the first range contributes a subleading $\hat c^{1/6}$, and the rest again is $\sim \hat c^{1/3}$, with a slightly different coefficient.

We have not taken tachyons into account. From our earlier discussion, with $n=0$ we have $8(g-1)$ of them. Unfortunately in this case we don't have a 5d supergravity description, and hence we don't know the volume of the tachyonic region to include in the count above. But the sum over $g$ converges rather quickly; the tachyonic contribution would have to grow very fast to spoil the power law result. In AdS$_7$ we have seen that tachyonic contributions actually worked the opposite way, taming an otherwise exponential behavior.

In the case with punctures, \eqref{eq:g-count} proceeds in a similar way. It would appear to lead to a power law only if $\kappa$ in \eqref{eq:S5} is taken to be $\gtrsim 4.62$, because of the remark below \eqref{eq:volMgn}. In that case, the count again goes as $\hat c^{1/3}$.
However, in this case we have more uncertainties about the final result than in \eqref{eq:g-count}, and we find it more likely that one of these tames the sum further, so that the choice of $\kappa$ should be irrelevant. 
\begin{itemize}
    \item[i)] The systolic suppression factor is only given in \cite{mirzakhani-petri} for $n=0$. But this effect did not modify much in \eqref{eq:g-count}.
    \item [ii)] When two punctures come too close, some new light states will appear from an M2 wrapping a segment on $\Sigma_g$ joining them; one would need to estimate the region in ${\mathcal M}_{g,n}$ where this does not happen. This suppression is probably also not very important.
    \item [iii)]  Many new vacua appear, corresponding to Higgsing each puncture. These are similar to the massive IIA vacua of AdS$_7$. 
    \item [iv)]
    Many new tachyonic modes appear. When punctures are maximal, they each contribute $5(N^2-1)$ tachyons, far in excess of the $8(g-1)$ we previously mentioned. Higgsing the punctures reduces this number, all the way to zero when they are closed. Our earlier discussion in this section showed that their structure is very similar to that we considered in AdS$_7$. It looks reasonable to us to assume that their effect is similar, and that eventually the effect of these tachyons counteracts the proliferation of vacua at the previous point.
\end{itemize}

All these uncertainties mean that more research would be needed to obtain a final estimate of the growth of class S vacua, but we still have evidence that it will grow with a power law in $\hat c$.

\subsection{\texorpdfstring{$\mathrm{AdS}_4$}{AdS4}}\label{sec:AdS4}
We consider M-theory on
\begin{align}
    \mathrm{AdS}_4 \times S^7/(\mathbb Z_{k_1}\times \mathbb Z_{k_2}\times \mathbb Z_{k_3}),
\end{align}
with $N$ units of $F_7$ on the $S^7/(\mathbb Z_{k_1}\times \mathbb Z_{k_2}\times \mathbb Z_{k_3})$. The gravitini are in irrep $\mathbf{8}_c$ of $Spin(8)$, which branches under $SU(4)\cong Spin(6) \subset Spin(8)$ as
\begin{align}
    \mathbf{8}_c \to \mathbf{6} + \mathbf{1} + \mathbf{1}.
\end{align}
We preserve supersymmetry by choosing
\begin{align}
    \mathbb Z_{k_1} \times \mathbb Z_{k_2} \times \mathbb Z_{k_3}\subset SU(4).
\end{align}
As in \autoref{sec:AdS5}, we can assume without loss of generality that
\begin{align}
    k_3 \mid k_2 \mid k_1.
\end{align}
Lastly, we assume that the weights associated to each discrete point $(N,k_1,k_2,k_3)$ is $V_{N,k_1,k_2,k_3}\sim 1$ as in \eqref{eq:weight-1}.

\paragraph{Scales.}
Let $R$ be the radius of $S^{7}$. Flux quantization and dimensional reduction give
\begin{align}
M_{11}^{6}\int_{S^{7}/(\mathbb Z_{k_{1}}\times\mathbb Z_{k_{2}}\times \mathbb Z_{k_3})}F_{7}\sim M_{11}^{6}\frac{R^{6}}{k_{1}k_{2}k_3}\sim N,\qquad
M_{4}^{2}\sim M_{11}^{9}\frac{R^{7}}{k_{1}k_{2}k_{3}}.
\end{align}
We get the dimensionless relations
\begin{align}
M_{11}R\sim (Nk_{1}k_{2}k_{3})^{1/6},\qquad
(M_{4}R)^{2}\sim N^{3/2}(k_{1}k_{2}k_{3})^{1/2},\qquad
\frac{M_{4}}{M_{11}}\sim N^{7/12}(k_{1}k_{2}k_{3})^{1/12}. \label{eq:AdS4-scales}
\end{align}

\paragraph{Towers.}
The $S^{7}$ KK tower sets
\begin{align}
\frac{m_{\rm KK}}{M_{4}}\sim\frac{1}{M_{4}R}\sim N^{-3/4}(k_{1}k_{2}k_{3})^{-1/4}\gtrsim\hat\mu. \label{eq:AdS7-KK}
\end{align}
We now check $p$-branes wrapping $q$-cycles. For $3\leq q \leq 5$, the minimal $q$-cycle volume scales as $R^{q}/(k_{1}k_{2}k_{3})$. Note that $p=5$ necessarily in this case since M2 branes can't wrap these cycles. Therefore
\begin{align}
    T_{5-q} \sim M_{11}^{6} \frac{R^q}{k_1k_2k_3}.
\end{align}
 In dimensionless terms, the tower scale is
\begin{align}
    \frac{T_{5-q}^{\frac 1 {6-q}}}{M_{4}} \sim \frac{M_{11}}{M_4} \left(\frac{(M_{11}R)^q}{k_1k_2k_3}\right)^{\frac 1 {6-q}}\sim N^{-\frac{7}{12}+\frac{q}{6(6-q)}}(k_1k_2k_3)^{-\frac{1}{4}} \gtrsim \hat\mu.
\end{align}
Comparing with \eqref{eq:AdS7-KK}, we see that the KK tower is lighter since $q\leq 5$.

For $q=2$, the smallest $2$-cycle is $R^2/(k_1k_2)$. Then
\begin{align}
    \frac{T_{p-2}^{\frac{1}{p-1}}}{M_4} \sim \frac{M_{11}}{M_4} \left(\frac{(M_{11}R)^2}{k_1k_2}\right)^{\frac{1}{p-1}} \sim N^{-\frac{7}{12}+\frac{1}{3(p-1)}}(k_1k_2)^{-\frac 1 {12}-\frac 2 {3(p-1)}}k_3^{-\frac{1}{12}+\frac{1}{3(p-1)}}\gtrsim\hat\mu.\label{eq:AdS7-q=2}
\end{align}
This inequality is potentially restricting. In fact, it turns out that $p=2$ corresponding to M2 wrapping a 2-cycle is among the minimal set of inequalities that imply all other inequalities as can be shown using techniques of \autoref{app:Motzkin}.

Lastly, we have the $q=1$ case 
\begin{align}
    \frac{T_{p-1}^{\frac{1}{p}}}{M_4} \sim \frac{M_{11}}{M_4} \left(\frac{M_{11}R}{k_1}\right)^{\frac{1}{p}} \sim N^{-\frac{7}{12}+\frac{1}{6p}}k_1^{-\frac{1}{12}-\frac{5}{6p}}(k_2k_3)^{-\frac 1 {12}+\frac 1 {6p}}\gtrsim\hat\mu.\label{eq:AdS7-q=1}
\end{align}
For $p=5$ it can be directly checked that this scale is heavier than the KK scale \eqref{eq:AdS7-KK}. For $p=2$, one needs to use the techniques of \autoref{app:Motzkin} to see it is implied by the other inequalities.

\paragraph{Orbifold choices and multiplicities.}
We will show that given $k$ for $\mathbb Z_{k}$, there are $\Psi_3(k)$ many $\mathbb Z_{k}\subset (\mathbb Z_{k})^3\subset U(1)^3$ choices, where $\Psi_n$ is the generalized Dedekind psi function as will be defined below.

Count the number of order $k$-elements in $(\mathbb Z_k)^3$ by enumerating all elements $(x,y,z)$ that are not all $p$ divisors for $p|k$. The ratio of the number of such elements to the whole group is $(1-1/p^3)$, and so the total number is
\begin{align}
    k^3 \prod_{p|k} \left( 1-\frac{1}{p^3} \right).
\end{align}
Divide it by the number of units in $\mathbb Z_{k}$ given by the Euler totient function $\varphi(k)$, so we get
$$\# \text{ of }\mathbb Z_{k} = \frac{k^3 \prod_{p|k} \left( 1-\frac{1}{p^3} \right)}{\varphi(k)} = k^2 \prod_{p|k} \left( 1+ \frac{1}{p} + \frac{1}{p^2} \right)=\Psi_3(k),$$
where $\Psi_n$ is the generalized Dedekind psi function 
\begin{align}
    \Psi_n(k) \equiv k^{n-1} \prod_{p|k} \sum_{i=0}^{n-1} p^{-i}.
\end{align}
The large $k$ behavior is given by \cite[Thm.~6.4]{mccarthy} 
\begin{align}
    \Psi_3(k) \sim k^2.
\end{align}
Therefore, the number of $\mathbb Z_{k_1}\times \mathbb Z_{k_2}\times \mathbb Z_{k_3}$ choices asymptotes as $(k_1k_2k_3)^2$. 
Similarly to before, we also have density measures associated to the divisors $k_3|k_2|k_1$ as $dk_3/k_3$ and $dk_2/k_2$.

\paragraph{Cylindrical decomposition.}
The integration region is defined by the bound on the KK tower scale \eqref{eq:AdS7-KK}
\begin{align}
    N^{-3/4}(k_{1}k_{2}k_3)^{-1/4}\gtrsim\hat\mu,
\end{align} 
and the tower from M2 wrapping $2$-cycles \eqref{eq:AdS7-q=2}
\begin{align}
    N^{-\frac 1 4}(k_1k_2)^{-\frac 3 4}k_3^{\frac 1 4}\gtrsim\hat\mu,
\end{align}
in addition to the positivity of the discrete data
\begin{align}
    N,k_1,k_2,k_3 \geq 1,
\end{align}
as well as divisor inequalities
\begin{align}
    k_1 \gtrsim k_2\gtrsim k_3.
\end{align}
All other bounds are implied by the bounds above. We use the built-in Mathematica \texttt{CylindricalDecomposition} function to put the inequalities in triangular form.

\paragraph{AdS count.}
The count is
\begin{align}\label{eq:AdS4-count-Q4}
\mathfrak N_{\mathrm{AdS}_{4}}^{(Q=4)}(\hat\mu)
\sim\int dNdk_1\frac{dk_{2}}{k_{2}}\frac{dk_3}{k_3}  k_{1}^2k_{2}^2 k_3^2\sim \hat\mu^{-28/5}.
\end{align}
In terms of the cosmological constant,
\begin{align}
    \sim |\hat\Lambda |^{-14/5}.
\end{align}
Using $(M_{4}R)^{2}\sim c$ and $M_{4}R\sim \hat\mu^{-1}$, we get the CFT count with central charge less than $\hat c$ as
\begin{align}
\mathfrak N_{\mathrm{CFT}_{3}}^{(Q=4)}(\hat c)\sim \hat c^{14/5}.
\end{align}
\paragraph{Special case ($Q=8$).}
Letting $k_3=1$ reduces to the special case $Q=8$ with the choice $\mathbb Z_{k_{1}}\times \mathbb Z_{k_2}\subset U(1)^2\subset Spin(4)$. We have $k_1k_2$ such choices. We get the counts
\begin{align}
\mathfrak N_{\mathrm{AdS}_{4}}^{(Q=8)}(\hat\mu)\sim \hat\mu^{-3},\\
\qquad
\mathfrak N_{\mathrm{CFT}_{3}}^{(Q=8)}(\hat c)\sim \hat c^{3/2}.
\end{align}
In terms of the cosmological constant,
\begin{align}
    \sim |\hat\Lambda |^{-3/2}
\end{align}

\paragraph{Special case ($Q=12$).}
Let $k_2=k_3=1$ and choose $\mathbb Z_{k_{1}}\subset  U(1)$ for which there is a unique choice given $k_1$. We get
\begin{align}
\mathfrak N_{\mathrm{AdS}_{4}}^{(Q=12)}(\hat\mu)&\sim \hat\mu^{-2},\\
\qquad
\mathfrak N_{\mathrm{CFT}_{3}}^{(Q=12)}(\hat c)&\sim \hat c.
\end{align}
In terms of the cosmological constant,
\begin{align}
    \sim |\hat\Lambda |^{-1}
\end{align}

\subsection{General discussion}
\label{sub:disc}

There are many more classes of known AdS$_d$ vacua in string and M-theory; we are aware that we have merely scratched the surface in this paper. An overview for $d\ge4$ is attempted in \cite[Chap.~11]{Tomasiello:2022dwe}; see in particular Tables 11.2--4 there.

A full analysis of all cases is beyond immediate reach mainly because of the analysis of tachyon regions. This requires having a lower-dimensional supergravity that includes all of a vacuum's tachyonic modes; as we illustrated in \autoref{sec:AdS7}, even when that is available, the complexity of the potential forces one to consider rough estimates rather than a complete analysis.

With this in mind, here are some general comments about the general count. We can divide known AdS vacua roughly into three categories.

\paragraph{Tame classes.} Some classes are characterized by a limited choice of internal spaces, and by several types of flux vacua that can be chosen freely. In (massive) IIA the internal space is a nearly-K\"ahler space, a coset, or a twistor space; all of these are currently known in small numbers. On each space, the cosmological constant is given as a product of powers, two typical behaviors being $\hat c\sim N^{3/2} k^{1/3}$ \cite{Aharony:2008ug} and $\hat c \sim N^{5/3}k^{1/3}$ \cite{Aharony:2010af,Guarino:2015jca}. A naive count of $\{N,k| N^i k^j < \hat c\}$ produces a result asymptotic to $\hat c^{1/\min(i,j)}$. 
    
\paragraph{Proliferation from internal topology.} In other classes, a possible proliferation comes from the number of possible internal spaces. The examples are currently AdS$_5\times {\rm SE}_5$ and AdS$_4\times {\rm SE}_7$. There are many known constructions of Sasaki--Einstein manifolds; each of these does lead to power laws. Focusing on IIB, in each class the central charge is of the form $N^2 f(k_i)$, with $f$ a homogeneous function of degree one of integer parameters $k_i$, $i=1,\ldots,n$ whose ratios obey a fixed bound; see e.g.~\cite[(3.35)]{Martelli:2005tp} for $Y^{p,q}$ and \cite[Sec.~8.1]{Collins:2015qsb} for Brieskorn--Pham singularities, both with $n=2$. A rough count taking into account the KK constraint alone leads to a growth $\sim \hat c^n$. Of course this would need to be reassessed as further methods of producing such spaces emerge. Reading this backwards, if the power-law scaling holds in general, then the number of SE spaces with volume above a given cutoff cannot be too large, and in particular not infinite.
    
\paragraph{Proliferation from internal branes.} The most insidious challenge comes from cases where many configurations are possible for internal branes. We saw two examples: AdS$_7$ in massive IIA (\autoref{sec:AdS7}), and class S AdS$_5$ in IIB (\autoref{sub:s}). Two more classes of this type are known:  
\begin{itemize}
    \item AdS$_6$ solutions in IIB \cite{DHoker:2016ujz,DHoker:2017mds,DHoker:2017zwj}. The internal space is a round $S^2$ fibered over a disc, with punctures at its boundary. In the original description \cite{DHoker:2016ujz,DHoker:2017mds}, these punctures describe NS5 $(p,q)$-branes, giving rise to several $SU(N_i)$ gauge groups. It is possible to Higgs the latter, a process whereby the five-branes migrate inside the disc and acquire seven-brane charges \cite{DHoker:2017zwj}, in a Myers-like effect clearly reminiscent of the above AdS$_7$ and AdS$_5$ cases.
    \item AdS$_4$ solutions in IIB with $\mathcal{N}=4$ \cite{DHoker:2007zhm,Assel:2011xz}. These are dual to the famous Hanany--Witten SCFT$_3$'s \cite{Hanany:1996ie}. The internal space is an $S^2\times S^2$ fibered over a disc, again with punctures at its boundary, this time representing either D5- or NS5-branes.
    While this picture is not commonly considered, it should be possible to view these solutions as originated from a solution with only two D3-brane stacks,\footnote{The solution with D3-branes is the Janus solution, which has two AdS$_5\times S_5$ asymptotics \cite{DHoker:2007zhm}.} with a Myers effect puffing one stack into D5-branes, and an S-dual Myers puffing up the other D3 stack into NS5-branes. 
\end{itemize}

Finding solutions of this type, where internal branes can be placed in many different configurations, is made very challenging by the branes' backreaction. It is possible, even likely, that more classes of this type exist, beyond the four we have just discussed (AdS$_7$ in IIA; class S AdS$_5$, AdS$_6$, AdS$_4$ $\mathcal{N}=4$ in IIB). This could be analyzed in the probe approximation. Even some apparently tame classes might actually hide a similar proliferation, upon closer inspection.

However, given the phenomena we described for AdS$_7$ solutions in \autoref{sec:AdS7}, it is natural to conjecture that a full treatment of  such apparently wild classes of solutions would be tamed by taking into account the volumes of tachyonic field theory spaces.

\section{Anthropic principle implications}\label{sec:anthropic}

We showed that the scaling relation
\begin{equation}
\mathfrak N(\Lambda) \sim |\Lambda|^{-b(d,\mathcal Q)} ,\qquad \Lambda\to 0,
\end{equation}
emerges as a universal feature of the AdS landscape once both massless and tachyonic directions are properly integrated over.  
Here \(b(d,\mathcal Q)\) is $O(1)$ and depends on the AdS dimension \(d\) and the amount of preserved supersymmetry~\(\mathcal Q\).  
This result replaces the exponential proliferation of vacua which we saw in the naive count for some examples with a more careful calculation leading to controlled polynomial growth.  
In this section we would like to discuss how this may be relevant for the anthropic principle.

The central anthropic argument for cosmological constant, first formulated by Weinberg~\cite{Weinberg:1987dv, Weinberg:1988cp}, asserts that the observed $\Lambda_{\text{obs}}>0$ must be small enough to allow the formation of galaxies and our existence, but no more fine-tuned than that.   Of course for this to work, there must be solutions with approximately constant $\Lambda$, and presumably many of them.  The question of how this distribution should look like to give a probabilistic interpretation of the observed value of the cosmological constant, has not been settled.

 It is natural to ask what our work about AdS count may suggest for positive values of $\Lambda$.  If we extrapolate from our AdS analysis to dS case, our analysis suggests such a derivation of the prior from AdS counting.  This continuation transforms our AdS counting result into a concrete statistical prediction about de~Sitter vacua--or more precisely quasi-dS vacua as dS conjectures \cite{Obied:2018sgi, Ooguri:2018wrx, Garg:2018reu} would lead us to believe, and thereby about the possible values of the cosmological constant or slowly evolving dark energy in our Universe.  
The `probability density' of vacua would be given by 
\begin{equation}
p(\Lambda) d\Lambda \propto \frac{d\mathfrak N}{d\Lambda} d\Lambda \sim \Lambda^{-b-1}d\Lambda,
\end{equation}
so that the measure of vacua grows polynomially as $\Lambda\!\to\!0$.  
The power-law divergence may ensure that the landscape is dense enough near the flat limit to support anthropic selection, but not so dense to overwhelm it.  Of course how such a count translates to picking a particular value of cosmological constant is part of the measure problem of the anthropic proposal \cite{PhysRevLett.74.846, Garriga:2005ee}, about which we have nothing to say. It would be interesting to see how this polynomial growth will impact the viability of the anthropic principle.

If, by contrast, one counts only the discrete set of vacua and ignores the continuous tachyonic directions, as we saw the number of AdS vacua would grow exponentially as in \eqref{eq:exp-Lambda-count}
\begin{equation}
\mathfrak{N}_{\text{naive}}(\Lambda) \sim \exp\!\big(C\,\Lambda^{-b}\big), \qquad C>0,
\end{equation}
in striking resemblance to the early proposals of Hawking~\cite{Hawking:1983hj,Hawking:1984hk}.  
In the Euclidean formulation of quantum cosmology, the semiclassical wavefunction of the universe behaves as, in $d=4$,
\begin{equation}
P_{\text{HH}}(\Lambda) \sim \exp\!\left(\frac{3\pi}{G\Lambda}\right),
\end{equation}
which favors $\Lambda\!\to\!0^+$ infinitely strongly. 
It is natural to interpret the de~Sitter entropy 
\(S_{\text{dS}} =3\pi/(G\Lambda)\)~\cite{Gibbons:1977mu} 
as a count of microstates. We are led to an entropy-weighted distribution
\begin{equation}
\mathfrak{N}_{\text{dS}}(\Lambda) \sim e^{S_{\text{dS}}} \sim \exp\!\left(\frac{3\pi}{G\Lambda}\right),
\end{equation}
an explicit exponential of the same form as our naive $\exp(C\Lambda^{-b})$ growth we got for some of the AdS cases.  In fact the spirit of this count for dS is similar to the naive count we got for AdS because it is the tree level contribution to the entropy and ignores the massless or light modes which in the more refined AdS computation softened the exponential growth.   
In both the Hartle--Hawking and de Sitter entropy pictures, such exponential weightings produce a non-normalizable measure dominated by infinitesimally small $\Lambda$.  
Our refined counting suggests a resolution to this pathology: including the volume of light modes may suppress the exponential proliferation and replace it with a polynomial law.  The exponential weighting $\exp(\Lambda^{-a})$ may be dynamically softened to a power law $\Lambda^{-b}$ leading potentially to non-zero but small values of $\Lambda$ as preferred values.\footnote{This would suggest that many of the $\Lambda\ll 1$ theories must have a large number $N$ of light modes with $N\gtrsim S_{dS}\sim 1/{\Lambda}$ for the volume factor to suppress the dS entropy.  It would be interesting to explore whether in the case of the extremal black holes given by AdS$_2$ a similar cancellation idea could relate the microscopic states of black hole to the area of the horizon.}  This also leads to a potential explanation of why we have few light moduli in our universe: as we saw in the AdS case, theories with many light moduli have exponentially small volume in the theory measure space and the ones contributing to measure are thus the ones with small number of moduli.

\acknowledgments

We thank T.~Grimm and N.~Mekareeya for useful discussions.   We would also like to thank the SCGP for the Simons Summer Physics Workshop 2025 for providing a productive research environment where this project was initiated.

The work of ZKB and CV is supported in part by a grant from the Simons Foundation (602883, CV) and the DellaPietra Foundation. AT is supported in part by the INFN, and by the MUR-PRIN contract 2022YZ5BA2.

\appendix
\section{Counts with instanton constraints}\label{app:instanton}

In the main text we imposed only the requirement that every potential tower satisfy $m_{\rm tower}> \mu$. This is the minimal condition for EFT control. We did not require that Euclidean wrapped brane instantons have large action. However, from the geometric viewpoint there are independent reasons to impose
\begin{align}
S_{\rm inst}\sim \frac{(M_{D}R)^{p+1}}{\prod_i k_i} \gtrsim 1 ,
\end{align}
for any Euclidean $p$-brane wrapped on an internal cycle. When this inequality fails, strong instanton corrections render the geometric description unreliable. The main text avoids imposing this inequality because even when the bulk geometric instanton action is small, the dual CFT can remain well-defined. The only essential requirement for the count is that all light towers lie above the cutoff.

Here we redo all counts under the conservative assumption that instanton inequalities are imposed in addition to the tower bounds. For the intermediate steps of the calculations see the supplementary Mathematica notebook.

\subsection{\texorpdfstring{$\mathrm{AdS}_7$}{AdS7}}

In the 11d case, instantons arise from Euclidean M2 branes on 3-cycles of $S^4/\mathbb Z_k$:
\begin{align}
S_{\rm M2} &\sim \frac{(M_{11}R)^{3}}{k}\sim N\gtrsim 1.
\end{align}
It turns out that the instanton inequality is always satisfied. So the result does not change.

\subsection{\texorpdfstring{$\mathrm{AdS}_5$}{AdS5}}

We consider IIB on $\mathrm{AdS}_5\times S^5/(\mathbb Z_{k_1}\times \mathbb Z_{k_2})$ as in \autoref{sec:AdS5-sphere}. Instantons arise from $p$-branes wrapping cycles on the orbifold:
\begin{align}\label{eq:AdS5-inst}
S_{\rm inst}&\sim \frac{(M_{10}R)^{p+1}}{k_1k_2}\sim N^{\frac{p+1}{4}} (k_1k_2)^{\frac{p+1}{4}-1} \gtrsim 1.
\end{align}

\begin{figure}
    \centering
    \includegraphics[width=0.4\linewidth]{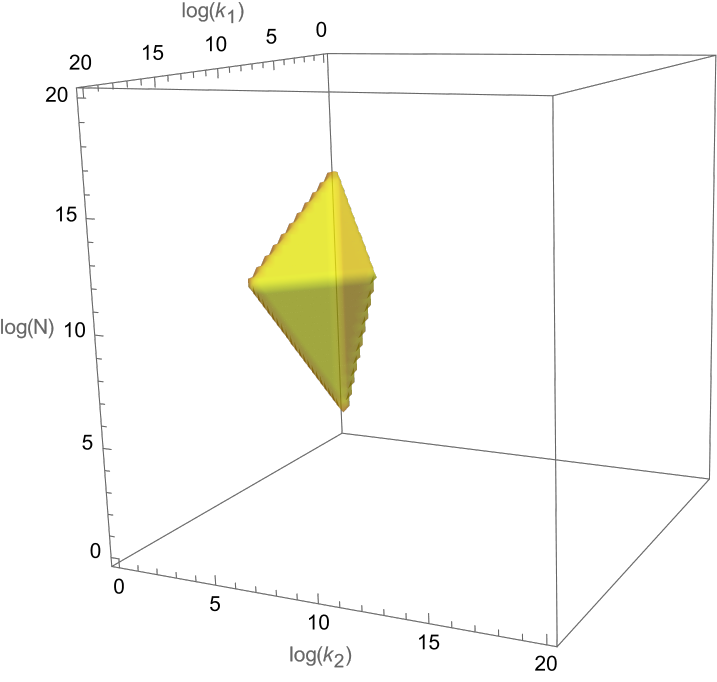}
    \caption{The integration region in $(\log(k_1),\log(k_2),\log(N))$ variables for the number of theories with tower scales above the cutoff $\hat\mu$ and instatons $S_{\rm inst}\gtrsim 1$ obtained from \eqref{eq:AdS5-ine1}, \eqref{eq:AdS5-ine3}, \eqref{eq:AdS5-ine4}, \eqref{eq:AdS5-inst} with $\log\hat\mu =-10$. Compare with \autoref{fig:ads5-region}.}
    \label{fig:ads5-instanton}
\end{figure}

It turns out that the wrapped F1 instanton constraint ($p=1$) is strictly stronger than the F1 tower inequality \eqref{eq:AdS5-F1} as well as the other instanton constraints.
Cylindrical decomposition with the instanton bound included yields a smaller allowed region in $(N,k_1,k_2)$ space as shown in \autoref{fig:ads5-instanton}.

As a result, the count becomes
\begin{align}
\mathfrak N^{(Q=4)}_{\text{AdS}_5}(\hat\mu)\sim \hat\mu^{-3}.
\end{align}
Compare with $\hat\mu^{-9/2}$ of \eqref{eq:AdS5-count-Q4}.

Fixing $k_2=1$ and changing the measure to $dk_1$ corresponds to the special case with $Q=8$. The instanton bound modifies only an $O(1)$ coefficient and does not change the power dependence:
\begin{align}
\mathfrak N^{(Q=8)}_{\text{AdS}_5}(\hat\mu)\sim \hat\mu^{-2}.
\end{align}

\subsection{\texorpdfstring{$\mathrm{AdS}_4$}{AdS4}}

We consider M-theory on $\mathrm{AdS}_4\times S^7/(\mathbb Z_{k_1}\times\mathbb Z_{k_2}\times\mathbb Z_{k_3})$.

In the main text the dominant wrapped-brane tower was the M2 on a $2$-cycle, corresponding to \eqref{eq:AdS7-q=2} for $p=2$.

The instanton actions are
\begin{align}
S_{\rm M2} &\sim \frac{(M_{11}R)^{3}}{k_1k_2k_3}\sim N^{-1/2}(k_1 k_2k_3)^{-1/2},\qquad
S_{\rm M5} \sim \frac{(M_{11}R)^{6}}{k_1k_2k_3}\sim N\gtrsim 1.
\end{align}
We see that the M5 instanton inequality is trivially satisfied. It turns out by techniques in \autoref{app:Motzkin} that the wrapped M2 instanton action gives an inequality stronger than the M2 on 2-cycle bound.

For $Q=4$ the count changes as
\begin{align}
\mathfrak N^{(Q=4)}_{\text{AdS}_4}(\hat\mu)\sim \hat\mu^{-4}.
\end{align}
Compare with $\hat\mu^{-28/5}$ of \eqref{eq:AdS4-count-Q4}. For $Q=8$ and $Q=12$ the counts change only up to a multiplicative $O(1)$ factor.

\section{Grassmannian geometry}\label{app:Grassmannian}
\subsection{Total volume}\label{app:tot-vol}

The goal of this appendix is to provide a detailed derivation and discussion of the volume of
arithmetic quotients of non-compact Grassmannians of orthogonal type,
\begin{align}
\widehat{\mathcal{M}}_{p,q}  =  O(p,q;\mathbb{Z}) \backslash O(p,q;\mathbb{R}) / (O(p)\times O(q)).
\end{align}
These spaces appear naturally as the scalar manifolds of half-maximal and minimal
supergravity theories obtained from M-theory compactifications on
$S^4/\mathbb{Z}_k$ and on massive IIA on $M_3$, as discussed in \autoref{sub:vol} of the main text.
In particular, for the case of AdS$_7$ vacua, the scalar manifold of the seven-dimensional
gauged supergravity truncation is
\begin{align}
\mathcal{M}_{\text{scal}} = \frac{O(3,\,2k^2)}{O(3)\times O(2k^2)}.
\end{align}
The quotient by $O(3,2k^2;\mathbb{Z})$ reflects the presence of discrete dualities (analogous
to U-dualities in M-theory), which render the total volume of moduli space finite.
This arithmetic quotient is crucial in the counting, because without such
quotienting the naive volume of the moduli space diverges exponentially with the distance
in field space. Scalar manifolds of this type also appear for supergravity models in other dimensions, for vector multiplets coupled to a gravity multiplet. Hence the present discussion will be relevant for a more complete treatment of class S theories (\autoref{sub:s}) or for other theories with a large proliferation of branes (AdS$_6$ and AdS$_4$ $\mathcal{N}=4$ in IIB, \autoref{sub:disc}).

Consider the symmetric space
\begin{align}
\mathcal M_{p,q} = \frac{O(p,q;\mathbb{R})}{O(p)\times O(q)},
\end{align}
which parametrizes $p$-dimensional spacelike planes in $\mathbb{R}^{p,q}$ equipped with the
quadratic form
\begin{align}
Q(x) = x_1^2 + \dots + x_p^2 - x_{p+1}^2 - \dots - x_{p+q}^2.
\end{align}
This space has real dimension $pq$, and negative curvature of order one in Planck units.
The scalar kinetic term in supergravity compactifications is proportional to the
$O(p,q)$-invariant metric on this space,
\begin{align}\label{eq:S_kin}
S_{\text{kin}} \,=\, \frac{1}{8}\int d^d x \sqrt{-g}\, \mathrm{Tr}(\partial_\mu M^{-1}\partial^\mu M),
\end{align}
where $M\in O(p,q)$ encodes the scalar degrees of freedom.  The measure induced by this metric
defines the local volume element on $\mathcal M_{p,q}$.

To fix conventions, take a coset representative \(V(x)\in O(p,q)\) defined up to right multiplication by \(O(p)\times O(q)\). Let
\begin{align}
V^{-1}\partial_\mu V  =  Q_\mu+P_\mu, \qquad 
Q_\mu\in\mathfrak{o}(p)\oplus\mathfrak{o}(q),\quad P_\mu\in\mathfrak{p},
\end{align}
be the Cartan decomposition, equivalently \(P_\mu=\tfrac{1-\theta}{2}(V^{-1}\partial_\mu V)\) with Cartan involution \(\theta\).
The \(O(p,q)\)-invariant metric on the symmetric space is
\begin{align}
ds^2 = -\tfrac12\,\mathrm{Tr}(P_\mu P^\mu).
\end{align}
Choosing the standard symmetric coset matrix
\begin{align}
M \equiv V^{T}\eta V, \qquad \eta=\mathrm{diag}(\underbrace{+1,\ldots,+1}_{p},\underbrace{-1,\ldots,-1}_{q}),
\end{align}
one finds the identity
\begin{align}
\mathrm{Tr}\!\left(\partial_\mu M^{-1}\partial^\mu M\right) = -4\,\mathrm{Tr}(P_\mu P^\mu).
\end{align}
Therefore the kinetic term \(\tfrac18\,\mathrm{Tr}(\partial_\mu M^{-1}\partial^\mu M)\) uses exactly the Maurer–Cartan-induced metric on \(O(p,q)/(O(p)\times O(q))\), so the associated measure coincides with the canonical Riemannian volume form on the symmetric space.

However, the full moduli space by itself has infinite volume unless it is further quotiented by a discrete arithmetic subgroup
$\Gamma\subset O(p,q;\mathbb{Z})$ in order to account for possible duality identifications.
The resulting arithmetic quotient
\begin{align}
\widehat{\mathcal{M}}_{p,q} = \Gamma \backslash \mathcal M_{p,q}
\end{align}
would then have finite total volume, and would be the appropriate physical moduli space to integrate over
in the AdS vacuum count.

The problem of computing $\mathrm{Vol}(\widehat{\mathcal{M}}_{p,q})$ is therefore the problem of computing
the covolume of $\Gamma$ inside $O(p,q;\mathbb{R})$.  This problem is classically solved
by the \emph{Siegel mass formula} for orthogonal groups.

Let $L$ be a non-degenerate integral lattice of signature $(p,q)$.
The arithmetic group $\Gamma = O(L,\mathbb{Z})$ consists of automorphisms of $L$ preserving
the quadratic form.
The \emph{genus} of $L$ is the set of all lattices locally isomorphic to $L$ over $\mathbb{Z}_{p'}$
for all primes $p'$ and over $\mathbb{R}$.
The \emph{mass} of the genus is defined as \cite{Siegel}
\begin{align}
m(L) = \sum_{[M]\in \mathrm{genus}(L)} \frac{1}{|\mathrm{Aut}(M)|},
\end{align}
and it measures the ``number of arithmetic lattices’’ of a given signature and determinant,
weighted by their automorphism groups.

The Siegel mass formula relates $m(L)$ to products of zeta and gamma factors:
for even unimodular lattices of rank $2d=p+q$ we have \cite{Moore:2015bba, BELOLIPETSKY2005221}
\begin{align}
m(L) = 2(d-1)! \frac{\zeta(d)}{(2\pi)^d}
\prod_{j=1}^{d-1} \frac{|B_{2j}|}{4j},
\end{align}
where $B_{2j}$ are Bernoulli numbers and $\zeta$ is the Riemann zeta function.
Odd unimodular lattices $I^{p,q}$ have similar expressions differing only by a power of~2.
This factor $m(L)$ accounts for the arithmetic contribution to the total moduli-space volume.

The volume of the arithmetic quotient $\widehat{\mathcal{M}}_{p,q}$ is obtained by multiplying the ratio of
orthogonal-group volumes with the genus mass $m(L)$:
\begin{align}
\mathrm{Vol}(\widehat{\mathcal{M}}_{p,q})
= \frac{\sigma(p+q)}{\sigma(p)\,\sigma(q)}\, m(L),
\end{align}
where
\begin{align}
\sigma(n) =
2^{\frac{n+1}{2}} \prod_{j=1}^n
\frac{(2\pi)^{(j+1)/2}}{\Gamma\!\left(\frac{j+1}{2}\right)}.
\end{align}
The factor $\sigma(n)$ is the volume of the orthogonal group $O(n)$ under the Haar measure
normalized so that the compact quotient $O(n+1)/O(n)$ has unit curvature radius. Intuitively, this expression separates the geometric and arithmetic data:
the ratio $\sigma(p+q)/(\sigma(p)\sigma(q))$ is the continuous geometric volume
of the noncompact symmetric space $X_{p,q}$, while $m(L)$ encodes the discrete lattice effects.

For our application, we are primarily interested in the case $p=3$ and $q=2k^2$,
which corresponds to the scalar manifold of seven-dimensional supergravity arising from
AdS$_7\times S^4/\mathbb{Z}_k$.  
We now analyze the asymptotic scaling of $\mathrm{Vol}(\widehat{\mathcal{M}}_{3,2k^2})$ for large $k$.

To study the asymptotic behavior, we approximate the factorial products using the
Stirling and Barnes $G$ asymptotics.  Letting $p=3$ and $q=2k^2$,
\begin{align}
\frac{\sigma(q+3)}{\sigma(3)\sigma(q)}
\propto
\frac{(2\pi)^{3q/2}}{
\Gamma\big(\tfrac{q}{2}+1\big)\,
\Gamma\big(\tfrac{q}{2}+\tfrac32\big)\,
\Gamma\big(\tfrac{q}{2}+2\big)},
\end{align}
and inserting the Bernoulli number product from the Siegel mass formula \cite{BELOLIPETSKY2005221},
\begin{align}
\prod_{j=1}^{(q+1)/2}\frac{|B_{2j}|}{4j}
=
\prod_{j=1}^{(q+1)/2}
\frac{\zeta(2j)}{(2\pi)^{2j}}\,(2j-1)!
\propto
(2\pi)^{-\frac{(q+1)(q+3)}{4}}
\!\prod_{j=1}^{(q+1)/2}(2j-1)!,
\end{align}
we use the Legendre duplication formula
\begin{align}
    \Gamma(2z)= (2\pi)^{-1/2}2^{2z-1/2}\Gamma(z) \Gamma(z+1/2),
\end{align}
to write
\begin{align}
\prod_{j=1}^{(q+1)/2}(2j-1)! \propto (2\pi)^{-(q+1)/4}2^{q(q+3)/4} G\left( \frac{q}{2} + \frac{3}{2} \right) G\left( \frac{q}{2} +2 \right).
\end{align}
Collecting all factors gives an expression of the form
\begin{align}
\mathrm{Vol}(\widehat{\mathcal{M}}_{3,q})
\sim
\frac{2^{q(q+3)/4}\,(2\pi)^{-q(q+1)/4}}{
\Gamma(\tfrac{q}{2}+1)\Gamma(\tfrac{q}{2}+2)}
G\!\left(\tfrac{q}{2}+\tfrac32\right)
G\!\left(\tfrac{q}{2}+2\right).
\end{align}
For large argument $z$, the Barnes $G$-function satisfies
\begin{align}
G(z+1) \sim
\,z^{\frac{z^2}{2}-\frac{1}{12}}\,
e^{-\frac{3z^2}{4}}(2\pi)^{z/2}.
\end{align}
Using this, one obtains the large-$q$ asymptotic scaling
\begin{align}\label{eq:G-as}
\mathrm{Vol}(\widehat{\mathcal{M}}_{3,q}) \sim
\left(\frac{q}{2\pi}\right)^{\!q^2/4}
e^{O(q^2)}.
\end{align}
Thus the volume has superexponential dependence on the number $q$ of negative directions.
This is the key quantitative statement: even though $\widehat{\mathcal{M}}_{p,q}$ has finite volume,
that volume increases faster than any polynomial as $q$ grows.

For our seven-dimensional case, we have $q=2k^2$, so that the dimension of the scalar manifold is $6k^2$.
Inserting into the asymptotic scaling gives
\begin{align}
\mathrm{Vol}(\widehat{\mathcal{M}}_{3,2k^2})
\sim
\left(\frac{k^2}{\pi}\right)^{\!k^4}
e^{O(k^2)}.
\end{align}
Up to constants, this can be summarized as
\begin{align}
\boxed{
\mathrm{Vol}(\widehat{\mathcal{M}}_{3,2k^2}) \sim k^{\,k^4}.
}
\end{align}
Hence, as $k$ increases, the scalar manifold volume grows faster than any power of~$k$.
This justifies the statement in the main text that taking $V_{N,k} = \mathrm{Vol}(\widehat{\mathcal{M}}_{3,2k^2})$
as the weight for each $(N,k)$ sector would ``grossly overcount'' the number of vacua.

For completeness, we note that odd unimodular lattices $I^{p,q}$,
which exist for all signatures $(p,q)$,
yield similar Siegel-mass prefactors that are approximately
\begin{align}
m(I^{p,q})
\sim 2^{d/2}\!\prod_{j=1}^{d-1}\!\frac{|B_{2j}|}{4j},
\end{align}
for odd $d$ and
\begin{align}
m(I^{p,q})
\sim 2^{d}\frac{(\frac{d}{2}-1)!}{(2\pi)^{d/2}}
\!\prod_{j=1}^{d-1}\!\frac{|B_{2j}|}{4j}
\end{align}
for even $d$ up to omitted constants. Inserting these into the volume formula modifies only subleading powers of $q$;
the asymptotic growth remains superexponential in $q$.
Therefore, the result
\begin{align}
\mathrm{Vol}(\widehat{\mathcal{M}}_{p,q}) \sim q^{\,q^2/4}
\end{align}
is universal across all unimodular lattice choices.

From the geometric standpoint, $\mathcal{M}_{p,q}$ is the moduli space of spacelike
$p$-planes in $\mathbb{R}^{p,q}$, and the arithmetic quotient ensures compactness at infinity.
As $q$ grows, the number of directions of negative curvature increases,
and the volume of the corresponding Grassmannian explodes superexponentially.
This mirrors the intuition from hyperbolic geometry:
the number of independent degrees of freedom in the metric (or scalar matrix $M$)
grows quadratically with the number of fields.

The physical consequence of this result is that the total scalar manifold volume
is dominated by a superexponential factor $\sim k^{k^4}$, far exceeding any polynomial growth.
As a result, if one were to integrate over the full arithmetic quotient when counting vacua,
the result would diverge catastrophically. This motivates restricting to a finite-radius
geodesic ball
so that only the physically accessible region of theory space contributes to the count.
 
\subsection{Ball volume}
We follow Helgason's classic textbook on symmetric spaces \cite{Helgason:GA}.

We want compute the invariant volume of a geodesic ball $B_R^{(p,q)}$ of radius $R$ in the non-compact Riemannian symmetric space
\begin{align}
\mathcal M_{p,q} = \frac{O(p,q)}{O(p)\times O(q)} ,
\end{align}
equipped with the metric induced from the Killing form on $\mathfrak{so}(p,q)$. This space arises as the scalar manifold of our theory in $\mathrm{AdS}_7$ and its local geometry controls the measure factor in counting vacua.

Let $\mathfrak g = \mathfrak{so}(p,q)$ with Cartan decomposition $\mathfrak g = \mathfrak k \oplus \mathfrak p$, where
\begin{align}
\mathfrak k = \mathfrak{so}(p)\oplus \mathfrak{so}(q), \qquad
\mathfrak p \simeq \mathrm{Hom}(\mathbb R^p,\mathbb R^q) .
\end{align}
Let $\mathfrak a\subset\mathfrak p$ denote a maximal abelian subspace, of dimension $d = \min(p,q)$, with elements written as
\begin{align}
H = \operatorname{diag}(r_1,\ldots,r_d,0,\ldots,0), \qquad r_i \in \mathbb R.
\end{align}
Every element $x\in\mathcal M_{p,q}$ admits a Cartan (polar) decomposition 
\begin{align}
x = k_1 e^H k_2 , \qquad k_1,k_2 \in O(p)\times O(q),\ H\in\mathfrak a^+,
\end{align}
where $\mathfrak a^+$ is the positive Weyl chamber $\{r_1\ge r_2\ge\cdots\ge r_d\ge0\}$.

In general, the integration in polar decomposition for a group $G$ with compact subgroup $K$ can be written as \cite[p. 271]{Helgason:GA}
\begin{align}\label{eq:polar_measure}
\int_{G} f(g)\,dg
 = \int_{K} \int_{K} \int_{\mathfrak{a}^+}
 f(k_1 e^Hk_2)\,J(H)\,dk_1\,dH\,dk_2 ,
\end{align}
where $J(H)$ is the exponential Jacobian and takes the general root–theoretic form
\begin{align}\label{eq:Jacobian_general}
J(H) = \prod_{\alpha\in\Sigma^+} (\sinh \alpha(H))^{m_\alpha},
\end{align}
where $\Sigma^+$ is the set of positive restricted roots of $(\mathfrak g,\mathfrak a)$ and $m_\alpha$ their multiplicities. For $\mathfrak g=\mathfrak{so}(p,q)$ we have
\begin{align}\label{eq:Jexplicit}
J(r)
 = \prod_{i=1}^d \sinh^{\,q-p}\! r_i
   \prod_{i<j}\sinh(r_i-r_j)\,\sinh(r_i+r_j).
\end{align}

To pass from the group integral \eqref{eq:polar_measure} to an integration formula on the symmetric space $G/K$ 
\begin{align}
    \int_{G/K} F(gK) d\mu,
\end{align}
one must account for the fact that the Cartan parametrization $g = k_1 e^H k_2$ becomes redundant after projection to $G/K$. The redundancy is controlled by the subgroup
\begin{align}
M := Z_K(A) = \{ m\in K \mid m e^H = e^H m \ \text{for all}\ H\in\mathfrak a \},
\end{align}
the centralizer of $A=\exp\mathfrak a$ in $K$. Since $m\in M$ fixes the base point $o = eK$ and commutes with $e^H$, we have
\begin{align}
(k m)e^H K = k m e^H K = k e^H m K = k e^H K,
\end{align}
so for fixed $H\in\mathfrak a^+$ the map $k\mapsto k e^H K$ depends only on the coset $kM$. Thus the true angular variable on the symmetric space is an element of $K/M$, not of $K$. This is the symmetric-space analogue of the familiar fact that in Euclidean space the angular coordinates live in $SO(n)/SO(n-1)\cong S^{n-1}$ rather than in $SO(n)$.

To make this precise, let $\widetilde F(g)=F(gK)$ be the right–$K$-invariant lift of $F$ to $G$. Using \eqref{eq:polar_measure},
\begin{align}
\int_G \widetilde F(g)dg
= \int_K\int_{\mathfrak a^+}\int_K
\widetilde F(k_1 e^H k_2)J(H)\, dk_1dHdk_2.
\end{align}
Right-$K$-invariance implies $\widetilde F(k_1 e^H k_2)=F(k_1 e^H K)$, so the integrand is independent of $k_2$ and the $k_2$-integral contributes a factor of $\mathrm{Vol}(K)$. Dividing by $\mathrm{Vol}(K)$ yields
\begin{align}
\int_{G/K} F(x)d\mu(x)
= C \int_K\int_{\mathfrak a^+} F(k e^H K)J(H)\,dkdH.
\end{align}

Next we decompose Haar measure on $K$ along the fibration $K\to K/M$. For any integrable $\phi:K\to\mathbb R$,
\begin{align}
\int_K \phi(k)dk
= \int_{K/M}\int_M \phi(km)\,dm\,d(kM).
\end{align}
Apply this to $\phi(k)=F(k e^H K)$, which is right-$M$-invariant. Then $\phi(km)=\phi(k)$ for all $m\in M$, and we obtain
\begin{align}
\int_K F(k e^H K)\,dk
= \mathrm{Vol}(M)\int_{K/M} F(k e^H K)\,d(kM).
\end{align}
Absorbing $\mathrm{Vol}(M)$ into the normalization constant gives the polar integration formula on the symmetric space
\begin{align}\label{eq:G/K-int}
\int_{G/K} F(x)\,d\mu(x)
= C' \int_{K/M}\int_{\mathfrak a^+}
F(k e^H K)J(H)\,dH\,d(kM).
\end{align}
Depending on normalization of the measure, one may set $C'=1$.

For $\mathcal M_{p,q}=O(p,q)/(O(p)\times O(q))$ one has $K=O(p)\times O(q)$ and
\begin{align}
M = O(p)\times O(q-p),
\end{align}
so
\begin{align}
K/M \cong \frac{O(p)\times O(q)}{O(p)\times O(q-p)}
\cong O(q)/O(q-p)
\cong V_{p,q},
\end{align}
the Stiefel manifold of orthonormal $p$-frames in $\mathbb R^q$. Thus the angular integration contributes a factor $\mathrm{Vol}(V_{p,q})$.

The invariant distance on $\mathcal M_{p,q}$ is obtained from the Killing metric restricted to $\mathfrak p$, and in the above coordinates the geodesic distance from the origin is
\begin{align}
R^2 = \sum_{i=1}^d r_i^2 .
\end{align}
We define the geodesic ball of radius $R$ as
\begin{align}
B_R^{(p,q)} = \{\,x\in\mathcal M_{p,q}\mid \sum_{i=1}^d r_i^2\le R\,\}.
\end{align}
Using the measure~\eqref{eq:G/K-int}, its volume can be written as
\begin{align}\label{eq:volBR_general}
\mathrm{Vol}(B_R^{(p,q)})
 = \mathrm{Vol}(V_{p,q}) \int_{\mathfrak a^+(R)} J(r)\,dr_1\cdots dr_r,
\end{align}
where $\mathrm{Vol}(V_{p,q})$ is the volume of the Stiefel manifold of orthonormal $p$-frames in $\mathbb R^{p+q}$ \cite{b1843332-5a23-3406-a810-5b0d74bb6a6d},
\begin{align}
\mathrm{Vol}(V_{p,q}) = \frac{2^p\pi^{p q/2}}{\Gamma_p(q/2)},
\end{align}
and $\mathfrak a^+(R)=\{r\in\mathfrak a^+\mid \sum_i r_i^2 \leq R^2\}$.

For small radius $R\ll1$, $\sinh r_i\approx  r_i$ and \eqref{eq:Jexplicit} reduces to a homogeneous polynomial of degree $pq-d$. Defining scaled coordinates $r_i = R s_i$ with $s_i\in[0,1]$, the integral in~\eqref{eq:volBR_general} gives
\begin{align}\label{eq:smallR}
\mathrm{Vol}(B_R^{(p,q)})
 &\approx \frac{2^p\pi^{pq/2}}{\Gamma_p(q/2)}\,R^{pq}
   \!\!\int_{1\ge s_1\ge\cdots\ge s_r\ge0}
   \!\!\!\!\!\!ds_1\cdots ds_r\,
   \prod_{i=1}^r s_i^{\,q-p}
   \prod_{i<j}(s_i^2-s_j^2).
\end{align}
The remaining integral is $O(q^{-p})$ for large $q$, so that asymptotically
\begin{align}\label{eq:ball_asymptotic}
\boxed{
\mathrm{Vol}(B_R^{(p,q)})
 \sim
 \frac{2^p\pi^{pq/2}}{\Gamma_p(q/2)}\,
 R^{pq}\,q^{-p},
 \qquad R\ll1 .
}
\end{align}
Thus, for fixed radius $R$ and $p$, the volume of a ball in $\mathcal M_{p,q}$ decreases superexponentially with the dimension $q$, consistent with the familiar $d^{-d/2}$ behavior of high-dimensional Euclidean balls.

\section{Motzkin Transposition Theorem}\label{app:Motzkin}

In several parts of the count, particularly in the analysis of the $T_{p-q}$ brane towers in
\autoref{sec:lowSUSY}, one encounters families of inequalities of the form
\begin{equation}
N^{-c_i} k^{-a_i} \gtrsim \hat\mu , \qquad i=1,\dots,m ,
\end{equation}
that restrict the allowed parameter space of flux integers $(N,k)$ for a fixed cutoff~$\hat\mu$, where $k$ may stand for a combination of $k_1,k_2,k_3$.
After taking logarithms,
\begin{equation}
c_i \log N + a_i \log k \lesssim -\log \hat\mu ,
\end{equation}
these become linear inequalities in the variables
\begin{align}
x=(\log N,\log k)^T .
\end{align}
The collection of all such inequalities defines a convex polyhedral region in the $(\log N,\log k)$
plane.  
Some inequalities may be redundant, in the sense that they are implied by the others.
To systematically identify and remove these redundancies, we use the
\emph{Motzkin Transposition Theorem}, a refinement of Farkas' lemma that characterizes linear
implications among inequalities.

The inequality form of Farkas’ lemma is as follows:
for any matrix $A$ and vector $b$, exactly one of the following holds:
\begin{align}
  &\text{(i)}\quad \exists\,x\ge0:  A x \le b, \\
  &\text{(ii)}\quad \exists\,y\ge0:   y^T A \ge 0,  y^T b < 0,
\end{align}
where $x\ge 0$ if all of its components are non-negative.
If (i) fails, then (ii) supplies a vector $y$, which defines a separating hyperplane between $b$ and the cone $\{Ax\mid x\geq 0\}$.  
The Motzkin Transposition Theorem \cite[Theorem 22.3]{Rockafellar1970} extends this to implications among inequalities.

\begin{theorem}[Motzkin Transposition Theorem]
Let $A$ be an $m\times n$ matrix, $a\in\mathbb{R}^m$,
$c\in\mathbb{R}^n$, and $b\in\mathbb{R}$.
Then
\begin{equation}
A x \le a  \Longrightarrow  c^T x \le b
\end{equation}
if and only if there exists $\lambda\in\mathbb{R}^m$ with $\lambda\ge0$
such that
\begin{equation}
c^T = \lambda^T A , \qquad \lambda^T a \le b .
\label{eq:motzkin-main}
\end{equation}
\end{theorem}

\noindent
The proof follows by applying the inequality form of Farkas to the infeasibility of
the system $\{ A x \le a ,\, c^T x > b\}$.
The vector $\lambda$ in~\eqref{eq:motzkin-main} plays the role of a
\emph{Farkas certificate}:
it gives the linear combination of the original inequalities sufficient to derive the target
inequality.

Each inequality $A_i x\le a_i$ defines a half-space in $\mathbb{R}^n$,
and their intersection forms a convex cone.
If the inequality $c^T x\le b$ is implied by these, then its bounding
hyperplane lies outside or tangent to the cone.
The vector $\lambda$ in~\eqref{eq:motzkin-main} specifies a convex combination
of the normal vectors $\{A_i\}$ that reproduces the normal $c$ of the outer hyperplane.
The inequalities
\begin{align}
c^T x = \lambda^T A x \le \lambda^T a \le b
\end{align}
make this chain explicit.

In our setup, all cutoff conditions and brane-tower bounds are of monomial type
\begin{equation}
N^{-c_i} k^{-a_i} \gtrsim \hat\mu \quad\Longleftrightarrow\quad
c_i \log N + a_i \log k \lesssim - \log\hat\mu .
\end{equation}
Let us denote by $A$ the matrix with rows $(c_i,a_i)$
and by $x=(\log N,\log k)^T$.
Different towers correspond to different rows of $A$.
If one tower’s bound is weaker than a convex combination of others,
the Motzkin theorem guarantees the existence of a nonnegative vector
$\lambda$ satisfying
\begin{equation}
(c_j,a_j) = \lambda^T A ,\qquad
\lambda^T a \lesssim b_j ,
\end{equation}
showing that the $j$th inequality is redundant.
We then drop it from the set without changing the feasible region.

As an example, we do the AdS$_5$ sphere quotient inequalities. In log form, the inequality from KK \eqref{eq:AdS5-KK} is
\begin{align}\label{eq:log-KK-AdS5}
    \frac 2 3 \log N + \frac 1 3 \log k_1 + \frac 1 3 \log k_2 \lesssim -\log \hat\mu.
\end{align}
The inequality from wrapped $q=1$ cycles \eqref{eq:AdS5-F1} is
\begin{align}\label{eq:log-AdS5F1}
    \left(\frac{5}{12} -\frac{1}{4p}\right) \log N + \left(\frac{1}{12} -\frac{1}{4p}\right) \log k_1 + \left(\frac{1}{12} +\frac{3}{4p}\right) \log k_2  \lesssim -\log \hat\mu. 
\end{align}
In addition, we have positivity bounds as
\begin{align}\label{eq:log-pos-AdS5}
    \log N, \log k_1,\log k_2 >0,
\end{align}
and also
\begin{align}
    \log k_1 - \log k_2 \lesssim 0.
\end{align}
We now show that $p=1$ in \eqref{eq:log-AdS5F1} together with the KK and positivity inequalities
\begin{align}
    Ax=\left(\begin{array}{ccc}
        \frac 2 3 & \frac 1 3 & \frac 1 3\\
        \frac 1 6 & -\frac 1 6 & \frac 5 6\\
        -1 & 0 & 0\\
        0 & -1 & 0\\
        0 & 0 & -1\\
        0 & 1 & -1
    \end{array}\right) \begin{pmatrix}\log N\\
    \log k_1\\
    \log k_2\end{pmatrix} \lesssim \begin{pmatrix}-\log \hat\mu\\
    -\log \hat\mu\\
    0\\
    0\\
    0\\
    0\end{pmatrix}.
\end{align}
 imply the $p=5$ inequality for \eqref{eq:log-AdS5F1}:
 \begin{align}
    \frac{11}{30}\log N +\frac{1}{30} \log k_1+\frac{7}{30} \log k_2 \lesssim -\log \hat\mu.
\end{align}
So we want to find the minimum
\begin{align}
    \underset{\lambda\geq 0}{\mathrm{argmin}}\,   \lambda^T \begin{pmatrix} -\log \hat\mu\\
    -\log \hat\mu\\
    0\\
    0\\
    0\\
    0\end{pmatrix}
\end{align}
with the restriction
\begin{align}
    \lambda^T \left(\begin{array}{ccc}
        \frac 2 3 & \frac 1 3 & \frac 1 3\\
        \frac 1 6 & -\frac 1 6 & \frac 5 6\\
        -1 & 0 & 0\\
        0 & -1 & 0\\
        0 & 0 & -1\\
        0 & 1 & -1
    \end{array}\right) = \begin{pmatrix}
        \frac{11}{30} & \frac{1}{30} & \frac{7}{30}
    \end{pmatrix}.
\end{align}
The solution can be found as
\begin{align}
    \lambda^T = \frac 1 {15}\begin{pmatrix}
        8 & 1 & 0 & 2 & 0 & 0
    \end{pmatrix}.
\end{align}
We therefore get the inequality
\begin{align}
    \lambda^T A x = \frac{11}{30}\log N + \frac{1}{30} \log k_1 + \frac{7}{30} \log k_2 \lesssim \lambda^T \begin{pmatrix}-\log \hat\mu\\
    -\log \hat\mu\\
    0\\
    0\\
    0\\
    0\end{pmatrix} = - \frac 3 5 \log \hat\mu< -\log \hat\mu.
\end{align}
This shows that the inequality for $p=5$ and $q=1$ in \eqref{eq:AdS5-F1} is implied by that for $p=1$ in addition to KK and positivity inequalities.

Conversely, we can check that the inequalities \eqref{eq:log-KK-AdS5}, \eqref{eq:log-AdS5F1}, and \eqref{eq:log-pos-AdS5} do not imply each other as the minimum value of $\lambda^T a$ is greater than $b$. This means this set of inequalities is irreducible.

\bibliographystyle{JHEP}
\bibliography{biblio.bib}

@article{Grimm:2021vpn,
    author = "Grimm, Thomas W.",
    title = "{Taming the landscape of effective theories}",
    eprint = "2112.08383",
    archivePrefix = "arXiv",
    primaryClass = "hep-th",
    doi = "10.1007/JHEP11(2022)003",
    journal = "JHEP",
    volume = "11",
    pages = "003",
    year = "2022"
}

@article{Grimm:2025lip,
    author = "Grimm, Thomas W. and Prieto, David and van Vliet, Mick",
    title = "{Tame embeddings, volume growth, and complexity of moduli spaces}",
    eprint = "2503.15601",
    archivePrefix = "arXiv",
    primaryClass = "hep-th",
    doi = "10.1103/d51c-j1s9",
    journal = "Phys. Rev. D",
    volume = "112",
    number = "10",
    pages = "106015",
    year = "2025"
}

@article{Hamada:2021yxy,
    author = "Hamada, Yuta and Montero, Miguel and Vafa, Cumrun and Valenzuela, Irene",
    title = "{Finiteness and the swampland}",
    eprint = "2111.00015",
    archivePrefix = "arXiv",
    primaryClass = "hep-th",
    doi = "10.1088/1751-8121/ac6404",
    journal = "J. Phys. A",
    volume = "55",
    number = "22",
    pages = "224005",
    year = "2022"
}

@article{Acharya:2006zw,
    author = "Acharya, Bobby Samir and Douglas, Michael R",
    title = "{A Finite landscape?}",
    eprint = "hep-th/0606212",
    archivePrefix = "arXiv",
    reportNumber = "IC-2006-42",
    month = "6",
    year = "2006"
}

@article{Siegel,
 ISSN = {0003486X, 19398980},
 URL = {http://www.jstor.org/stable/1969191},
 author = {Carl Ludwig Siegel},
 journal = {Annals of Mathematics},
 number = {3},
 pages = {577--622},
 publisher = {[Annals of Mathematics, Trustees of Princeton University on Behalf of the Annals of Mathematics, Mathematics Department, Princeton University]},
 title = {On the Theory of Indefinite Quadratic Forms},
 urldate = {2025-11-09},
 volume = {45},
 year = {1944}
}

@article{Belin:2025qjm,
    author = "Belin, Alexandre and Maloney, Alexander and Seefeld, Florian",
    title = "{A measure on the space of CFTs and pure 3D gravity}",
    eprint = "2509.04554",
    archivePrefix = "arXiv",
    primaryClass = "hep-th",
    month = "9",
    year = "2025"
}

@article{Moore:2015bba,
    author = "Moore, Gregory W.",
    title = "{Computation Of Some Zamolodchikov Volumes, With An Application}",
    eprint = "1508.05612",
    archivePrefix = "arXiv",
    primaryClass = "hep-th",
    month = "8",
    year = "2015"
}

@article{Ooguri:2002gx,
    author = "Ooguri, Hirosi and Vafa, Cumrun",
    title = "{World sheet derivation of a large $N$ duality}",
    eprint = "hep-th/0205297",
    archivePrefix = "arXiv",
    reportNumber = "CALT-68-2386, CITUSC-02-019, HUTP-02-A018",
    doi = "10.1016/S0550-3213(02)00620-X",
    journal = "Nucl. Phys. B",
    volume = "641",
    pages = "3--34",
    year = "2002"
}

@article{b1843332-5a23-3406-a810-5b0d74bb6a6d,
 ISSN = {07492170},
 URL = {http://www.jstor.org/stable/4355803},
 author = {Yasuko Chikuse},
 journal = {Lecture Notes-Monograph Series},
 pages = {177--193},
 publisher = {Institute of Mathematical Statistics},
 title = {Invariant Measures on Stiefel Manifolds with Applications to Multivariate Analysis},
 urldate = {2025-11-09},
 volume = {24},
 year = {1994}
}

@article{Garriga:2005ee,
    author = "Garriga, Jaume and Vilenkin, Alexander",
    editor = "Sasaki, Misao and Soda, Jiro and Tanaka, Takahiro",
    title = "{Anthropic prediction for Lambda and the Q catastrophe}",
    eprint = "hep-th/0508005",
    archivePrefix = "arXiv",
    doi = "10.1143/PTPS.163.245",
    journal = "Prog. Theor. Phys. Suppl.",
    volume = "163",
    pages = "245--257",
    year = "2006"
}

@article{PhysRevLett.74.846,
  title = {Predictions from Quantum Cosmology},
  author = {Vilenkin, Alexander},
  journal = {Phys. Rev. Lett.},
  volume = {74},
  issue = {6},
  pages = {846--849},
  numpages = {0},
  year = {1995},
  month = {Feb},
  publisher = {American Physical Society},
  doi = {10.1103/PhysRevLett.74.846},
  url = {https://link.aps.org/doi/10.1103/PhysRevLett.74.846}
}

@article{Obied:2018sgi,
    author = "Obied, Georges and Ooguri, Hirosi and Spodyneiko, Lev and Vafa, Cumrun",
    title = "{De Sitter Space and the Swampland}",
    eprint = "1806.08362",
    archivePrefix = "arXiv",
    primaryClass = "hep-th",
    reportNumber = "CALT-TH-2018-020, IPMU18-0100",
    month = "6",
    year = "2018"
}

@article{Garg:2018reu,
    author = "Garg, Sumit K. and Krishnan, Chethan",
    title = "{Bounds on Slow Roll and the de Sitter Swampland}",
    eprint = "1807.05193",
    archivePrefix = "arXiv",
    primaryClass = "hep-th",
    doi = "10.1007/JHEP11(2019)075",
    journal = "JHEP",
    volume = "11",
    pages = "075",
    year = "2019"
}

@book{Rockafellar1970,
  author    = {Rockafellar, R. Tyrrell},
  title     = {Convex Analysis},
  publisher = {Princeton University Press},
  address   = {Princeton, NJ},
  year      = {1970},
  series    = {Princeton Landmarks in Mathematics},
  isbn      = {978-0-691-01586-6}
}

@article{Ooguri:2018wrx,
    author = "Ooguri, Hirosi and Palti, Eran and Shiu, Gary and Vafa, Cumrun",
    title = "{Distance and de Sitter Conjectures on the Swampland}",
    eprint = "1810.05506",
    archivePrefix = "arXiv",
    primaryClass = "hep-th",
    doi = "10.1016/j.physletb.2018.11.018",
    journal = "Phys. Lett. B",
    volume = "788",
    pages = "180--184",
    year = "2019"
}

@book{Helgason:GA,
  author    = {Sigurdur Helgason},
  title     = {Geometric Analysis on Symmetric Spaces},
  series    = {Mathematical Surveys and Monographs},
  volume    = {39},
  publisher = {American Mathematical Society},
  address   = {Providence, RI},
  year      = {1994},
  isbn      = {978-0-8218-1538-0},
  doi       = {10.1090/surv/039},
  note      = {Second printing with corrections}
}

@article{BELOLIPETSKY2005221,
title = {The mass of unimodular lattices},
journal = {Journal of Number Theory},
volume = {114},
number = {2},
pages = {221-237},
year = {2005},
issn = {0022-314X},
doi = {https://doi.org/10.1016/j.jnt.2005.02.007},
url = {https://www.sciencedirect.com/science/article/pii/S0022314X05000569},
author = {Mikhail Belolipetsky and Wee Teck Gan}
}

@article{Gibbons:1977mu,
    author = "Gibbons, G. W. and Hawking, S. W.",
    title = "{Cosmological Event Horizons, Thermodynamics, and Particle Creation}",
    doi = "10.1103/PhysRevD.15.2738",
    journal = "Phys. Rev. D",
    volume = "15",
    pages = "2738--2751",
    year = "1977"
}

@article{Hawking:1983hj,
    author = "Hawking, S. W.",
    editor = "Fang, Li-Zhi and Ruffini, R.",
    title = "{The Quantum State of the Universe}",
    reportNumber = "PRINT-84-0117 (CAMBRIDGE)",
    doi = "10.1016/0550-3213(84)90093-2",
    journal = "Nucl. Phys. B",
    volume = "239",
    pages = "257",
    year = "1984"
}

@article{Hawking:1984hk,
    author = "Hawking, S. W.",
    title = "{The Cosmological Constant Is Probably Zero}",
    reportNumber = "Print-84-0116 (CAMBRIDGE)",
    doi = "10.1016/0370-2693(84)91370-4",
    journal = "Phys. Lett. B",
    volume = "134",
    pages = "403",
    year = "1984"
}

@article{Weinberg:1988cp,
    author = "Weinberg, Steven",
    editor = "Hsu, Jong-Ping and Fine, D.",
    title = "{The Cosmological Constant Problem}",
    reportNumber = "UTTG-12-88",
    doi = "10.1103/RevModPhys.61.1",
    journal = "Rev. Mod. Phys.",
    volume = "61",
    pages = "1--23",
    year = "1989"
}

@article{Weinberg:1987dv,
    author = "Weinberg, Steven",
    title = "{Anthropic Bound on the Cosmological Constant}",
    reportNumber = "UTTG-06-87",
    doi = "10.1103/PhysRevLett.59.2607",
    journal = "Phys. Rev. Lett.",
    volume = "59",
    pages = "2607",
    year = "1987"
}

@article{DeLuca:2018zbi,
    author = "De Luca, G. Bruno and Gnecchi, Alessandra and Lo Monaco, Gabriele and Tomasiello, Alessandro",
    title = "{Holographic duals of 6d RG flows}",
    eprint = "1810.10013",
    archivePrefix = "arXiv",
    primaryClass = "hep-th",
    reportNumber = "CERN-TH-2018-229",
    doi = "10.1007/JHEP03(2019)035",
    journal = "JHEP",
    volume = "03",
    pages = "035",
    year = "2019"
}

@article{Ooguri:2006in,
    author = "Ooguri, Hirosi and Vafa, Cumrun",
    title = "{On the Geometry of the String Landscape and the Swampland}",
    eprint = "hep-th/0605264",
    archivePrefix = "arXiv",
    reportNumber = "CALT-68-2600, HUTP-06-A017",
    doi = "10.1016/j.nuclphysb.2006.10.033",
    journal = "Nucl. Phys. B",
    volume = "766",
    pages = "21--33",
    year = "2007"
}

@article{Delgado:2024skw,
    author = "Delgado, Matilda and van de Heisteeg, Damian and Raman, Sanjay and Torres, Ethan and Vafa, Cumrun and Xu, Kai",
    title = "{Finiteness and the emergence of dualities}",
    eprint = "2412.03640",
    archivePrefix = "arXiv",
    primaryClass = "hep-th",
    reportNumber = "MPP-2024-224, CERN-TH-2024-204",
    doi = "10.21468/SciPostPhys.19.2.047",
    journal = "SciPost Phys.",
    volume = "19",
    number = "2",
    pages = "047",
    year = "2025"
}

@article{Vafa:2005ui,
    author = "Vafa, Cumrun",
    title = "{The String landscape and the swampland}",
    eprint = "hep-th/0509212",
    archivePrefix = "arXiv",
    reportNumber = "HUTP-05-A043",
    month = "9",
    year = "2005"
}

@article{Breitenlohner:1982bm,
    author = "Breitenlohner, Peter and Freedman, Daniel Z.",
    title = "{Positive Energy in anti-De Sitter Backgrounds and Gauged Extended Supergravity}",
    reportNumber = "PRINT-82-0420 (MIT)",
    doi = "10.1016/0370-2693(82)90643-8",
    journal = "Phys. Lett. B",
    volume = "115",
    pages = "197--201",
    year = "1982"
}

@article{Breitenlohner:1982jf,
    author = "Breitenlohner, Peter and Freedman, Daniel Z.",
    title = "{Stability in Gauged Extended Supergravity}",
    reportNumber = "Print-82-0500 (MIT)",
    doi = "10.1016/0003-4916(82)90116-6",
    journal = "Annals Phys.",
    volume = "144",
    pages = "249",
    year = "1982"
}

@article{DeLuca:2021ojx,
    author = "De Luca, G. Bruno and De Ponti, Nicol{\`o} and Mondino, Andrea and Tomasiello, Alessandro",
    title = "{Cheeger bounds on spin-two fields}",
    eprint = "2109.11560",
    archivePrefix = "arXiv",
    primaryClass = "hep-th",
    doi = "10.1007/JHEP12(2021)217",
    journal = "JHEP",
    volume = "12",
    pages = "217",
    year = "2021"
}

@article{Bhattacharya:2024tjw,
    author = "Bhattacharya, Ritabrata and Katyal, Abhay and Varela, Oscar",
    title = "{Class S superconformal indices from maximal supergravity}",
    eprint = "2411.16837",
    archivePrefix = "arXiv",
    primaryClass = "hep-th",
    doi = "10.1103/PhysRevLett.134.181601",
    journal = "Phys. Rev. Lett.",
    volume = "134",
    number = "18",
    pages = "181601",
    year = "2025"
}

@article{Gaiotto:2009gz,
    author = "Gaiotto, Davide and Maldacena, Juan",
    title = "{The gravity duals of ${\cal N}=2$ superconformal field theories}",
    eprint = "0904.4466",
    archivePrefix = "arXiv",
    primaryClass = "hep-th",
    doi = "10.1007/JHEP10(2012)189",
    journal = "JHEP",
    volume = "10",
    pages = "189",
    year = "2012"
}

@article{Tachikawa:2017aux,
    author = "Tachikawa, Yuji and Yonekura, Kazuya",
    title = "{Anomalies involving the space of couplings and the Zamolodchikov metric}",
    eprint = "1710.03934",
    archivePrefix = "arXiv",
    primaryClass = "hep-th",
    reportNumber = "IPMU-17-0140",
    doi = "10.1007/JHEP12(2017)140",
    journal = "JHEP",
    volume = "12",
    pages = "140",
    year = "2017"
}

@article{mirzakhani-zograf,
  title        = {Towards large genus asymptotics of intersection numbers on moduli spaces of curves},
  author       = {Mirzakhani, Maryam and Zograf, Peter},
  journal      = {Geometric and Functional Analysis},
  volume       = {25},
  number       = {4},
  pages        = {1258--1289},
  year         = {2015},
  doi          = {10.1007/s00039-015-0336-5},
  eprint       = {1112.1151},
    archivePrefix = {arXiv},
    primaryClass  = {math.AG}
}

@article{mirzakhani-petri,
  title        = {Lengths of closed geodesics on random surfaces of large genus},
  author       = {Mirzakhani, Maryam and Petri, Bram},
  journal      = {Commentarii Mathematici Helvetici},
  volume       = {94},
  number       = {4},
  pages        = {869--889},
  year         = {2019},
  doi          = {10.4171/CMH/477},
  eprint       = {1710.09727},
  archivePrefix = {arXiv},
  primaryClass  = {math.GT},
}

@article{Maldacena:2000mw,
    author = "Maldacena, Juan Martin and Nunez, Carlos",
    editor = "Duff, Michael J. and Liu, J. T. and Lu, J.",
    title = "{Supergravity description of field theories on curved manifolds and a no go theorem}",
    eprint = "hep-th/0007018",
    archivePrefix = "arXiv",
    doi = "10.1142/S0217751X01003937",
    journal = "Int. J. Mod. Phys. A",
    volume = "16",
    pages = "822--855",
    year = "2001"
}

@book{Tachikawa:2013kta,
    author = "Tachikawa, Yuji",
    title = "{${\cal N}=2$ supersymmetric dynamics for pedestrians}",
    eprint = "1312.2684",
    archivePrefix = "arXiv",
    primaryClass = "hep-th",
    reportNumber = "UT-13-42, IPMU-13-0234, UT-13-42, IPMU-13-0234",
    doi = "10.1007/978-3-319-08822-8",
    volume = "890",
    month = "12",
    year = "2013",
    publisher = {Springer},
}

@article{Cremonesi:2015bld,
    author = "Cremonesi, Stefano and Tomasiello, Alessandro",
    title = "{6d holographic anomaly match as a continuum limit}",
    eprint = "1512.02225",
    archivePrefix = "arXiv",
    primaryClass = "hep-th",
    doi = "10.1007/JHEP05(2016)031",
    journal = "JHEP",
    volume = "05",
    pages = "031",
    year = "2016"
}

@article{Apruzzi:2019ecr,
    author = "Apruzzi, Fabio and Bruno De Luca, G. and Gnecchi, Alessandra and Lo Monaco, Gabriele and Tomasiello, Alessandro",
    title = "{On AdS$_{7}$ stability}",
    eprint = "1912.13491",
    archivePrefix = "arXiv",
    primaryClass = "hep-th",
    reportNumber = "CERN-TH-2019-232",
    doi = "10.1007/JHEP07(2020)033",
    journal = "JHEP",
    volume = "07",
    pages = "033",
    year = "2020"
}

@article{Mekareeya:2016yal,
    author = "Mekareeya, Noppadol and Rudelius, Tom and Tomasiello, Alessandro",
    title = "{T-branes, Anomalies and Moduli Spaces in 6D SCFTs}",
    eprint = "1612.06399",
    archivePrefix = "arXiv",
    primaryClass = "hep-th",
    doi = "10.1007/JHEP10(2017)158",
    journal = "JHEP",
    volume = "10",
    pages = "158",
    year = "2017"
}

@article{Cordova:2016emh,
    author = "Cordova, Clay and Dumitrescu, Thomas T. and Intriligator, Kenneth",
    title = "{Multiplets of Superconformal Symmetry in Diverse Dimensions}",
    eprint = "1612.00809",
    archivePrefix = "arXiv",
    primaryClass = "hep-th",
    doi = "10.1007/JHEP03(2019)163",
    journal = "JHEP",
    volume = "03",
    pages = "163",
    year = "2019"
}

@article{Apruzzi:2013yva,
    author = "Apruzzi, Fabio and Fazzi, Marco and Rosa, Dario and Tomasiello, Alessandro",
    title = "{All AdS$_7$ solutions of type II supergravity}",
    eprint = "1309.2949",
    archivePrefix = "arXiv",
    primaryClass = "hep-th",
    doi = "10.1007/JHEP04(2014)064",
    journal = "JHEP",
    volume = "04",
    pages = "064",
    year = "2014"
}

@article{Apruzzi:2015wna,
    author = "Apruzzi, Fabio and Fazzi, Marco and Passias, Achilleas and Rota, Andrea and Tomasiello, Alessandro",
    title = "{Six-Dimensional Superconformal Theories and their Compactifications from Type IIA Supergravity}",
    eprint = "1502.06616",
    archivePrefix = "arXiv",
    primaryClass = "hep-th",
    doi = "10.1103/PhysRevLett.115.061601",
    journal = "Phys. Rev. Lett.",
    volume = "115",
    number = "6",
    pages = "061601",
    year = "2015"
}

@book{Tomasiello:2022dwe,
    author = "Tomasiello, Alessandro",
    title = "{Geometry of String Theory Compactifications}",
    doi = "10.1017/9781108635745",
    isbn = "978-1-108-63574-5, 978-1-108-47373-6",
    publisher = "Cambridge University Press",
    month = "1",
    year = "2022"
}

@article{DHoker:2016ujz,
    author = "D'Hoker, Eric and Gutperle, Michael and Karch, Andreas and Uhlemann, Christoph F.",
    title = "{Warped $AdS_6\times S^2$ in Type IIB supergravity I: Local solutions}",
    eprint = "1606.01254",
    archivePrefix = "arXiv",
    primaryClass = "hep-th",
    doi = "10.1007/JHEP08(2016)046",
    journal = "JHEP",
    volume = "08",
    pages = "046",
    year = "2016"
}

@article{DHoker:2017mds,
    author = "D'Hoker, Eric and Gutperle, Michael and Uhlemann, Christoph F.",
    title = "{Warped $AdS_6\times S^2$ in Type IIB supergravity II: Global solutions and five-brane webs}",
    eprint = "1703.08186",
    archivePrefix = "arXiv",
    primaryClass = "hep-th",
    doi = "10.1007/JHEP05(2017)131",
    journal = "JHEP",
    volume = "05",
    pages = "131",
    year = "2017"
}

@article{DHoker:2017zwj,
    author = "D'Hoker, Eric and Gutperle, Michael and Uhlemann, Christoph F.",
    title = "{Warped $AdS_6\times S^2$ in Type IIB supergravity III: Global solutions with seven-branes}",
    eprint = "1706.00433",
    archivePrefix = "arXiv",
    primaryClass = "hep-th",
    doi = "10.1007/JHEP11(2017)200",
    journal = "JHEP",
    volume = "11",
    pages = "200",
    year = "2017"
}

@article{Hanany:1996ie,
    author = "Hanany, Amihay and Witten, Edward",
    title = "{Type IIB superstrings, BPS monopoles, and three-dimensional gauge dynamics}",
    eprint = "hep-th/9611230",
    archivePrefix = "arXiv",
    reportNumber = "IASSNS-HEP-96-121",
    doi = "10.1016/S0550-3213(97)00157-0",
    journal = "Nucl. Phys. B",
    volume = "492",
    pages = "152--190",
    year = "1997"
}

@article{DHoker:2007zhm,
    author = "D'Hoker, Eric and Estes, John and Gutperle, Michael",
    title = "{Exact half-BPS Type IIB interface solutions. I. Local solution and supersymmetric Janus}",
    eprint = "0705.0022",
    archivePrefix = "arXiv",
    primaryClass = "hep-th",
    reportNumber = "UCLA-07-TEP-09",
    doi = "10.1088/1126-6708/2007/06/021",
    journal = "JHEP",
    volume = "06",
    pages = "021",
    year = "2007"
}

@article{Assel:2011xz,
    author = "Assel, Benjamin and Bachas, Costas and Estes, John and Gomis, Jaume",
    title = "{Holographic Duals of D=3 N=4 Superconformal Field Theories}",
    eprint = "1106.4253",
    archivePrefix = "arXiv",
    primaryClass = "hep-th",
    doi = "10.1007/JHEP08(2011)087",
    journal = "JHEP",
    volume = "08",
    pages = "087",
    year = "2011"
}

@article{Aharony:2008ug,
    author = "Aharony, Ofer and Bergman, Oren and Jafferis, Daniel Louis and Maldacena, Juan",
    title = "{N=6 superconformal Chern-Simons-matter theories, M2-branes and their gravity duals}",
    eprint = "0806.1218",
    archivePrefix = "arXiv",
    primaryClass = "hep-th",
    reportNumber = "WIS-12-08-JUN-DPP",
    doi = "10.1088/1126-6708/2008/10/091",
    journal = "JHEP",
    volume = "10",
    pages = "091",
    year = "2008"
}

@article{Aharony:2010af,
    author = "Aharony, Ofer and Jafferis, Daniel and Tomasiello, Alessandro and Zaffaroni, Alberto",
    title = "{Massive type IIA string theory cannot be strongly coupled}",
    eprint = "1007.2451",
    archivePrefix = "arXiv",
    primaryClass = "hep-th",
    doi = "10.1007/JHEP11(2010)047",
    journal = "JHEP",
    volume = "11",
    pages = "047",
    year = "2010"
}

@article{Guarino:2015jca,
    author = "Guarino, Adolfo and Jafferis, Daniel L. and Varela, Oscar",
    title = "{String Theory Origin of Dyonic N=8 Supergravity and Its Chern-Simons Duals}",
    eprint = "1504.08009",
    archivePrefix = "arXiv",
    primaryClass = "hep-th",
    reportNumber = "NIKHEF-2015-011",
    doi = "10.1103/PhysRevLett.115.091601",
    journal = "Phys. Rev. Lett.",
    volume = "115",
    number = "9",
    pages = "091601",
    year = "2015"
}

@book{montgomery-vaughan,
  title={Multiplicative number theory I: Classical theory},
  author={Montgomery, Hugh L and Vaughan, Robert C},
  series={Cambridge Studies in Advanced Mathematics},
  number={97},
  year={2007},
  publisher={Cambridge University Press}
}

@book{mccarthy,
  title={Introduction to arithmetical functions},
  author={McCarthy, Paul J},
  year={2012},
  publisher={Springer}
}

@article{Martelli:2005tp,
    author = "Martelli, D. and Sparks, J. and Yau, Shing-Tung",
    title = "{The Geometric dual of $a$-maximisation for Toric Sasaki--Einstein manifolds}",
    eprint = "hep-th/0503183",
    archivePrefix = "arXiv",
    reportNumber = "CERN-PH-TH-2005-047, HUTP-05-A0012",
    doi = "10.1007/s00220-006-0087-0",
    journal = "Commun. Math. Phys.",
    volume = "268",
    pages = "39--65",
    year = "2006"
}

@article{Collins:2015qsb,
    author = "Collins, Tristan C. and Sz{\'e}kelyhidi, G{\'a}bor",
    title = "{Sasaki-Einstein metrics and K-stability}",
    eprint = "1512.07213",
    archivePrefix = "arXiv",
    primaryClass = "math.DG",
    doi = "10.2140/gt.2019.23.1339",
    journal = "Geom. Topol.",
    volume = "23",
    pages = "1339--1413",
    year = "2019"
}

\end{document}